\documentclass[12pt]{article}

\usepackage{amsmath,amsfonts,amssymb,bbm}
\usepackage[mathscr]{eucal}
\usepackage[latin1]{inputenc}

\newcommand{\Tr}{\ensuremath{\textrm{Tr}}}
\newcommand{\TrN}{\ensuremath{\textrm{tr}_N}}
\newcommand{\dAslash}{D_A\!\!\!\!\!\!\!\slash\;\;}
\newcommand{\fslash}{\ensuremath{\sigma\cdot F}}
\newcommand{\psdo}{\ensuremath{\Psi\textrm{DO}}}
\newtheorem{prp}{Proposition}[section]
\newcommand{\Ds}{D\!\!\!\!\slash}

\begin{document}

\title{Yang-Mills action from minimally coupled bosons on 
  $\mathbbm R^4$ and on the $4$D Moyal plane}
\author{Juha Loikkanen and Cornelius Pauf\/ler}
\date{\normalsize
  Mathematical Physics/AlbaNova\\
  Royal Institute of Technology\\
  SE-10691 Stockholm\\
  Sweden\\[3ex]
  \large
  23 July 2004}

\maketitle

\thispagestyle{empty}

\begin{abstract}
\noindent We consider bosons on (Euclidean) $\mathbbm R^4$ that are
minimally coupled to an external Yang-Mills field. We compute the
logarithmically divergent part of the cut-off regularized quantum
effective action of this system.  We confirm the known result that
this term is proportional to the Yang-Mills action.\\ 
We use pseudodifferential operator methods throughout to prepare the ground
for a generalization of our calculation to the noncommutative
four-dimensional Moyal plane $\mathbbm R^4_\theta$. We also 
include a detailed comparison of our cut-off regularization to heat
kernel techniques.\\ 
In the case of the noncommutative space, we complement the usual
technique of asymptotic expansion in the momentum variable with
operator theoretic arguments in order to keep separated quantum from
noncommutativity effects. We show that the result from the commutative
space $\mathbbm R^4$ still holds if one replaces all pointwise
products by the noncommutative Moyal product.
\end{abstract}

\newpage
\section{Introduction}

In this article we study the determinant of certain differential
operators. Such determinants naturally arise in quantum field theory
at the one loop level.  As the determinant of an operator on an
infinite dimensional Hilbert space is not an a priori well-defined
object, one has to choose some regularization scheme. The latter means
generally the choice of a recipe for how to replace the formal
expressions by something that is both amenable to a rigorous
definition and close in its properties. In our case of the
regularization of determinants, a common starting point is the
well-known identity
\begin{equation}
  \log\det A=\Tr \log A
\end{equation}
which holds for (finite dimensional) matrices $A$. The task is now to
give meaning to the trace on the right hand side, since the operators
of interest do not have a finite trace in general. In this article, we
restrict the trace to run over a subspace of our Hilbert space
only. Loosely speaking, this subspace is spanned by wave functions
that have a momentum expectation value smaller than a certain cut-off
$\Lambda$. The precise definition will follow below.  It is known that
the cut-off regularized logarithm of the determinant, now viewed as a
function of $\Lambda$, contains a term that scales like $\log\Lambda$
for large $\Lambda$. This term is closely related to the Wodzicki
residue for the operator under consideration, a quantity that is of
interest in the study of infinite dimensional geometry, see
\cite{Paycha:2001} for a recent review.

Motivated by the observation \cite{Lang01} that for fermions minimally
coupled to an external Yang-Mills field, the logarithmically divergent
part of the cut-off regularized logarithm of the determinant of the
(massive) Dirac operator is proportional to the corresponding
Yang-Mills action, we consider the case of bosons in an external
Yang-Mills field on $\mathbbm R^4$. With our work, we confirmed that
the above result also applies to the bosonic case.  The former result
was proposed to be interpreted in two ways. On the one hand, the
spectral action principle \cite{SAP} states that the spectrum of the
Dirac operator should provide exhaustive information about the
complete bare action including the Yang-Mills expression. On the other
hand, it is known that the logarithmically divergent part plays a
critical role in the selection of the finite part of an effective
action because of its behavior under rescaling of the regularization
parameter $\Lambda$. From the latter viewpoint, it is desirable that
the logarithmically divergent term in the regularized effective action
produce expressions that occur in the complete bare action already.

To understand the connection between these two interpretations, it is
interesting to consider the case of (scalar) bosons coupled to an
external Yang-Mills field.

It is generally accepted that space-time might loose its smooth
properties at very small scales. One possible mathematical framework
for this is noncommutative geometry \cite{Connes}. We are interested
in a particular example, the four-dimensional Moyal plane
\cite{SpectralTriples}, also known as noncommutative flat space
$\mathbbm R^4_\theta$. Roughly speaking, the $4$D Moyal plane differs
from its Euclidean counterpart $\mathbbm R^4$ in that there is an
uncertainty relation for the simultaneous measurement of coordinates
coming from the non-vanishing commutator
\begin{equation*}
  [x^\mu,x^\nu]=i\Theta^{\mu\nu}.
\end{equation*}
Here $x^\mu$, $\mu=1,\ldots,4$ are coordinates of $\mathbbm R^4$ and
$\Theta$ is some (antisymmetric) matrix. In this article we take
$\Theta$ to be proportional to the constant symplectic matrix, see
Eg.~(\ref{defTheta}).  We refer to \cite{DFR} for a treatment of
Lorentz covariant generalization of this equation.

Although the results of our analysis for the case of bosons on the
commutative space $\mathbbm R^4$ are not new and can be found already
in deWitt's book \cite{deWitt}, our consistent use of
pseudodifferential operator methods technically makes possible the
generalization to the noncommutative Moyal plane. We refer to
\cite{Vass03} and \cite{Moyal} for a generalization of heat kernel
regularization calculations to the noncommutative torus and the Moyal
plane, respectively.

The remainder of this article is structured as follows. Section 2 sets
up the notation used in our work and states the results in the form of
two propositions.  Section 3 provides the necessary tools from the
theory of pseudodifferential operators.  The proofs of the statements
from Section 2 can be found in Sections 4 and 5, with detailed
calculations postponed to the appendix. Also, Section 4 contains
additional arguments that make contact with the case of fermions on
$\mathbbm R^4$ and to an alternative regularization scheme, the heat
kernel regularization. Section 6 concludes with what we consider to be
the lessons from our calculations.

\section{Notation and statement of the results}

We consider the Klein-Gordon operator with minimally coupled external
field on the four-dimensional flat Euclidean space $\mathbbm R^4$,
given by
\begin{equation}
  \begin{split}
    \Box_A&=D^\mu_A D^{\phantom{\mu}}_{A,\mu}=(\partial ^\mu+ i e
    A^\mu)(\partial _\mu + i e A_\mu)\\
    &=\partial^\mu\partial_\mu +i e \partial^\mu A_\mu +2 i e
    A^\mu\partial_\mu - e^2 A^\mu A_\mu\\
    &=\Box_0 +i e \partial ^\mu A_\mu + 2 i e A^\mu\partial_\mu
    -e^2 A^\mu A_\mu.
  \end{split}
\end{equation}
Here, $\mu=1,\ldots,4$ are the (Euclidean) indices of $\mathbbm R^4$,
$\partial_\mu=\frac{\partial}{\partial x^\mu}$, and $A_\mu$ are
$\textrm{gl}_N$-valued Yang-Mills fields on $\mathbbm R^4$.  The
bosonic wave functions are elements of the Hilbert space
\begin{equation*}
\mathcal H=L^2(\mathbbm R^4)\otimes \mathbbm C^N_{\textrm{color}},
\end{equation*}
where the last factor carries a $\textrm{gl}_N$ representation from
the external Yang-Mills fields.  As an unbounded operator in $\mathcal
H$, $\Box_0$ can be defined on smooth functions in $\mathcal H$ by its
formal expression and then extended to a self-adjoint operator. We
assume the Yang-Mills fields $A_\mu$ to be regular, i.e. to be smooth
and to fall off (together with all their derivatives) at infinity like
$|x|^{-2-\epsilon}$, $\epsilon>0$. The latter assumption ensures that
all our spatial integrals below will converge. Also, for regular
$A_\mu$, the self-adjoint extension of $\Box_A$ can be computed from
that of $\Box_0$. In what follows, we will not distinguish between the
formal expression for $\Box_A$ and its self-adjoint extension.

We consider the cut-off regularized logarithm of the determinant
of the massive Klein-Gordon operator
\begin{equation}\label{defdet}
  S_\Lambda(A)
  :=
  \Tr_\Lambda\left(\log(\frac{-\Box_A+m^2}{\Lambda_0^2})
  -\log(\frac{-\Box_0+m^2}{\Lambda_0^2})\right),
\end{equation}
where the cut-off regularized Hilbert space trace $\Tr_\Lambda$ sums
over states with momentum bounded by $\Lambda$. More precisely, if $D$
is an operator on $\mathcal H$ and $\Tr$ denotes the operator trace on
$\mathcal H$, then the cut-off trace is defined by
\begin{equation}\label{defcutoff}
\Tr_\Lambda D:= \Tr \big\{\theta(\Lambda^2+\Box_0) D\big\},
\end{equation}
where $\theta$ is the Heaviside step function.  As is well-known, the
expression (\ref{defdet}) occurs in Quantum Field Theory as the
one-loop effective action. Using the formal identity
$\log\det=\Tr\log$, it can be viewed as the generalization of the
determinant to operators on an infinite-dimensional Hilbert space. 

The parameter $\Lambda_0$ has been introduced to balance physical
dimensions.  It also provides a useful tool for cross checking since
in the result of our calculations, it should cancel.  Moreover, in the
definition of the regularized determinant, we have subtracted a term
containing the free Klein-Gordon operator as a reference. Whereas this
term is needed for turning the expression under the trace into a
pseudodifferential operator, it also comes in --- at least in the
corresponding expression for fermions --- when interpreting the
determinant as a subsummation of the one loop diagrams in the Feynman
path integral \cite{QFTbook}.

The regularized determinant (\ref{defdet}) has an asymptotic expansion
in $\Lambda$ for large values of $\Lambda$ as
\begin{equation}\label{S-asymptext}
\begin{split}
   S_\Lambda(A)
  &= c_2(A,m) \,\Lambda^2 +c_1(A,m) \,\Lambda^1
  +c_{\textrm{log}}(A,m) \,\log\Lambda
  +c_0(A,m) +\ldots,
\end{split}
\end{equation}
where the dots indicate terms that vanish at least like $\frac1\Lambda$
in the limit $\Lambda\rightarrow \infty$.

We are interested in the coefficient
$c_{\textrm{log}}(A,m)=c_{\textrm{log}}(A)$.
\begin{prp}\label{prp:comm}
  For the regularized determinant
  $S_\Lambda(A)$ defined as above,
  the coefficient $c_{\textrm{log}}(A,m)$ is proportional to the
  Yang-Mills action of $A_\mu$,
  \begin{equation}\label{clog}
    c_{\textrm{log}}(A)=\frac{1}{96\pi^2}\int_{\mathbbm R^4} d^4x\,
    \TrN (F^{\mu\nu} F_{\mu\nu}),
  \end{equation}
  where $\TrN $ is the matrix trace in $\textrm{gl}_N$ and
  the curvature $F_{\mu\nu}$ of $A_\mu$ is given by
  \begin{equation}
    F_{\mu\nu}=\partial_\mu A_\nu -\partial_\nu A_\mu
    +i e [A_\mu,A_\nu].
  \end{equation}
\end{prp}
The proof is contained in Section \ref{logdiv}. Note that the
numerical factor in front of $F^{\mu\nu}F_{\mu\nu}$ differs from the
one obtained in \cite[Eqs (24.16) {\sl et seq.}]{deWitt} by
$\frac12$. By the considerations below, this can be understood as
coming from the usage of a non-gauge invariant regularization for
$S_\Lambda(A)$. However, the latter allows a straight\-for\-ward
ge\-ne\-ra\-li\-za\-tion to the noncommutative Moyal plane.

\noindent
It is well-known that
\begin{equation}
  (\dAslash)^2=\mathbbm 1_4 \Box_A-ie \fslash,
\end{equation}
where $\dAslash=\gamma^\mu(\partial_\mu+ie A_\mu)$,
$\fslash=\frac14 \sigma^{\mu\nu} F_{\mu\nu}
=\frac12\gamma^\mu\gamma^\nu F_{\mu\nu}$, and $\gamma^\mu$
are the four-dimensional gamma matrices, i.e. $4\times4$ matrices that satisfy
\begin{equation}
\gamma^\mu\gamma^\nu+\gamma^\nu\gamma^\mu=2\eta^{\mu\nu}\mathbbm 1_4,
\end{equation}
$\eta^{\mu\nu}$ being the Euclidean flat metric.  Using this identity,
we are able to rederive the result of \cite{Lang01} concerning the
determinant of the Dirac operator.  This is demonstrated in Section
\ref{Fermion}.  Our computation has the advantage of avoiding
extensive calculations involving the gamma matrices $\gamma^\mu$.

\noindent
It is at first sight surprising that the non-gauge invariant
definition of the determinant yields a gauge invariant logarithmically
divergent part. It is therefore natural to consider the manifestly
gauge invariant expression
\begin{equation}\label{ginvdet-defn}
  \tilde S_\Lambda(A)
  :=
  \Tr_\Lambda^{\Box_A} \log\big(\frac{-\Box_A+m^2}{\Lambda^2_0}\big)-
  \Tr_\Lambda\log\big(\frac{-\Box_0+m^2}{\Lambda^2_0}\big),
\end{equation}
where in the first trace, the cut-off is taken w.r.t. the operator
$\Box_A$ rather than $\Box_0$. As before, $\tilde S_\Lambda(A)$
has an asymptotic expansion,
\begin{equation}\label{ginvdet-expansion}
  \begin{split}
    \tilde S_\Lambda(A)
    &= \tilde c_2(A,m) \,\Lambda^2 +\tilde c_1(A,m) \,\Lambda^1
    +\tilde c_{\textrm{log}}(A,m) \,\log\Lambda
    +\tilde c_0(A,m) +\ldots,
  \end{split}
\end{equation}
the dots subsuming terms scaling at least like $\frac1\Lambda$.  A
calculation in Section \ref{regdep} reveals that the coefficient
$\tilde c_{\textrm{log}}(A,m)=\tilde c_{\textrm{log}}(A)$ in
(\ref{ginvdet-expansion}) equals half of the corresponding expression
in $S_\Lambda(A)$,
\begin{equation}\label{res:regdep}
  \tilde c_{\textrm{log}}(A)=\frac12 c_{\textrm{log}}(A).
\end{equation}
This result agrees with the one obtained in \cite{deWitt}.

A widely used alternative regularization of the determinant of a
differential operator makes use of the $\zeta$-function and the
asymptotic expansion of the trace of the heat kernel operator.  We
want to compare our coefficient with earlier results that have been
obtained with these methods \cite{Vass-Manual,Elizalde}. In this
approach, one considers asymptotic expansions for the trace of the
heat operator,
\begin{equation}
  \begin{split}
    K(f,D)&:=\Tr_{L^2}(f e^{-tD})
    =
    t^{-2} a_0(f,D)+t^{-\frac32} a_1(f,D)+\ldots+\\
    &\hspace*{17em} +a_4(f,D)+\ldots,
  \end{split}
\end{equation}
for small $t$, where $f$ is some function on $\mathbbm R^4$ that
serves as a regulator for the spatial integrals. The rightmost dots
indicate terms that fall off at least linearly in $t$.  As the heat
trace has to be integrated on the positive axis together with the
function $t$, the logarithmically divergent contribution to the heat
kernel regularized trace is given by the coefficient $a_4(f,D)$. For
the comparison of this coefficient to our result, let
$c_{(\cdot)}(f,A)$ etc.\ be the coefficients in the expanssion of
$S_\Lambda(A)$, now spatially regularized in the same way as
$K(f,D)$. In Section \ref{heatkernel}, it is shown that the
coefficient $\tilde c_{\textrm{log}}(f,A)$ in the asymptotic expansion
of $\tilde S_\Lambda(A)$ differs from the corresponding expression
obtained via heat kernel regularization methods by a term proportional
to $m^4$,
\begin{equation}\label{heat-kernel-result}
  \textstyle
  -\tilde c_{\textrm{log}}(f,A) + \frac1{32\pi^2} m^4 \int_{\mathbbm R^4}
  d^4x \,f(x) = a_4(f,{\textstyle \frac{-\Box_A+m^2}{\Lambda^2_0}}).
\end{equation}
The additional mass term can be traced back to the usage of the
reference operator $-\Box_0+m^2$ in (\ref{defdet}). The calculations
using the heat operator can be generalized to the noncommutative 4D
torus \cite{Vass03} and the noncommutative Moyal plane
\cite{Moyal}. The only change one encounters is that in all
expressions, the commutative product of functions has to be replaced
by the noncommutative product $\star$.

The main part of our article is devoted to the study of the case of
the $4$D Moyal plane as the underlying (noncommutative) ``space''.
In this case, the algebra of functions on $\mathbbm R^4$ is furnished
with the (noncommutative) Moyal-Weyl product
$\star:=\star_\Theta$. The latter is defined by the integral formula
\begin{equation}\label{defmoyal}
  f\star g(x)=\frac{1}{(2\pi)^4}\int_{\mathbbm R^4}\int_{\mathbbm R^4}
  d^4y\,d^4\xi\,
  e^{i\,\xi(x-y)}\,f(x-{\textstyle
    \frac12}\Theta\xi)\,g(y),
\end{equation}
where $\Theta$ is a $4\times 4$ matrix defined by
\begin{equation}\label{defTheta}
  \Theta=\theta
  \left(\begin{array}{cc}0&\mathbbm 1_2\\
    -\mathbbm 1_2&0
  \end{array}
  \right).
\end{equation}
for the real parameter $\theta$. In our calculations we do not use
asymptotic expansions of this product in powers of $\theta$.

On the Moyal plane, we consider the generalized Klein-Gordon operator
\begin{equation}
  \Box_A^\theta=\partial^\mu\partial_\mu+i e(\partial^\mu A_\mu)\star
  +2 i e A^\mu\star \partial_\mu -e^2 (A^\mu\star A_\mu)\star,
\end{equation}
where $f\star$ is a shorthand notation for the operator that
$\star$-multiplies smooth wave functions in $\mathcal H$ from the left
by the (smooth) function $f$. We define $S^\theta_\Lambda(A)$ and
$c_{(\cdot)}^\theta(A,m)$ in analogy with the formulas (\ref{defdet})
and (\ref{S-asymptext}) above. Then, our main result is the following.
\begin{prp}\label{prp:star}
  For minimally coupled bosonic fields on the (non\-commu\-ta\-tive)
  $4$D Moyal plane, the above formula (\ref{clog}) holds with the
  commutative products replaced by the noncommutative Moyal-Weyl
  product, i.e. we have
 \begin{equation}
    c_{\textrm{log}}^\theta(A)=\frac{1}{96\pi^2}\int_{\mathbbm
    R^4}d^4x\,
    \TrN F^{\theta,\mu\nu}\star F^\theta_{\mu\nu},
  \end{equation}
  where
  \begin{equation}
    F^\theta_{\mu\nu}=\partial_\mu A_\mu-\partial_\nu A_\mu
    +e[A_\mu, A_\nu]_{\star}.
  \end{equation}
\end{prp}

\section{Pseudodifferential operator methods}

In our work, we deal with a restricted class of {\em
pseudodifferential operators} (\psdo) which suits our purposes. The
statements below may be found in Shubin's book \cite{Shubin}. We
consider \psdo s that act on smooth and compactly supported wave
functions $u$ as ($x=(x^\mu)$, $\mu=1,\ldots,4$, and likewise $y$, $p$
describe points in $\mathbbm R^4$; $x\,p=\sum_\mu x^\mu\, p^\mu$
denotes the scalar product, $|x|$ is the length of the vector $x$)
\begin{equation*}
(Au)(x)=\int_{\mathbbm R^4}\frac{d^4p}{(2\pi)^4}\,\int_{\mathbbm
    R^4}d^4y\; \sigma[A](x,p)\,u(y)\,e^{\textrm{i}\,p(x-y)},
\end{equation*}
where the {\em symbol} $\sigma[A]$ of $A$ is a smooth function
that allows an {\em asymptotic expansion} in $p$ according to
\begin{equation*}
\sigma[A](x,p)\sim\sum_{r=0}^\infty \, \sigma_{m-r}[A](x,p).
\end{equation*}
Here, $\sim$ means that for each $s$, the finite sum
$\sum_{r=0}^s\sigma_{m-r}[A]$ approximates $\sigma[A]$ up to a
function that falls off at most as $|p|^{m-(s+1)}$ for large $|p|$:
\begin{equation*}
|\partial_x^\alpha\partial_p^\beta\left(\sigma[A](x,p)
-\sum_{r=0}^s\sigma_{m-r}[A](x,p)\right)|\leq
 C_{\alpha\beta}(1+|p|^2)^{\frac{m-(s+1)-|\beta|}{2}}
\end{equation*}
for all multi-indices
$\alpha=(\alpha_1,\ldots,\alpha_4)$,
$\beta=(\beta_1,\ldots,\beta_4)$, where
\begin{equation*}
  \textstyle
  \partial_x^\alpha
  =\big(\frac{\partial}
  {\partial x^1}\big)^{\alpha_1}\cdots
  \big(\frac{\partial}{\partial x^4}\big)^{\alpha_4},
\end{equation*}
$|\alpha|=\alpha_1+\cdots+\alpha_4$, and $C_{\alpha\beta}$ are
constants. The number $m$ above is called the {\em order} of $A$. For a
given symbol, there are many different asymptotic expansions. One
particular choice is the asymptotic expansion in terms of {\em homogeneous
symbols} $\sigma_{m-r}^h[A]$, i.e. smooth functions that in addition
satisfy
\begin{equation*}
  \sigma_{m-r}^h[A](x,\lambda
  p)=\lambda^{m-r}\,\sigma_{m-r}^h[A](x,p),\textrm{ for }
  |p|=1,\;\lambda>1.
\end{equation*}
The first term $\sigma_m^h[A]$ in an asymptotic expansion in
homogeneous summands is termed the {\em principal symbol}.

An asymptotic expansion encodes the information of a given symbol
$\sigma[A]$ up to an additive function that falls off in $p$ like
a Schwartz test function. This piece of information will be
sufficient for our purposes.

While the expansion in homogeneous symbols is appropriate to
discuss invariant notions such as the residue of a \psdo, the
expansions obtained from recursion relations in the computation of
resolvents of operators are not of this type in general. The two
types, however, are related to each other through a finite
resummation at every order of the infinite sum.

The action of the \psdo{s} considered here can be extended to
smooth functions, leading to the useful formula
\begin{equation*}
\sigma[A](x,p)=e^{-\textrm{i} xp}\,A\,e^{\textrm{i} xp}.
\end{equation*}

For the product $AB$ of two \psdo s $A$ and $B$ with respective
symbols $\sigma[A]$ and $\sigma[B]$, one has the following asymptotic
expansion of the symbol:
\begin{equation}\label{psdo-product}
  \sigma[A]\ast\sigma[B](x,p)=\sigma[AB](x,p)
  \sim\sum_\alpha \frac{(-i)^{|\alpha|}}{\alpha!}
  \partial_p^\alpha\sigma[A](x,p)\,\partial_x^\alpha\sigma[B](x,p),
\end{equation}
where the sum runs over all $4$-indices $\alpha$ and we have used the
notation $\alpha!=\alpha_1!\cdots\alpha_{4}!$. We will use $\ast$
whenever we mean this product of symbols, in contrast to the
noncommutative product $\star$ defined later on.

Interpreting $A$ as an operator in the Hilbert space $L^2(\mathbbm
R^4)\otimes\mathbbm C^N$, we can compute the trace of $A$ from its
symbol according to
\begin{equation*}
\Tr (A)=\int_{\mathbbm R^4}\,\frac{d^4p}{(2\pi)^4}\,
\int_{\mathbbm R^4} d^4x\, \TrN\,\sigma[A](x,p),
\end{equation*}
where $\TrN$ denotes the matrix trace over the
$\textrm{gl}_N$-part of the symbol.

For operators $A$ that do not have a (finite) trace, one considers the
cut-off trace
\begin{equation*}
\Tr_\Lambda (A)=\int_{|p|\leq\Lambda}\,\frac{d^4p}{(2\pi)^4}\,
\int_{\mathbbm R^4} d^4x\, \TrN\,\sigma[A](x,p).
\end{equation*}
Clearly, this coincides with the previous definition of the cut-off
regularized trace, Eg.~(\ref{defcutoff}).

The above expression has an asymptotic expansion in $\Lambda$, as can
be seen from the asymptotic expansion of the symbol $\sigma[A]$ in
homogeneous symbols. In this case, there appears a term scaling like
$\log\Lambda$. On the other hand, the \emph{Wodzicki residue}
\cite{WodzRes} of the operator $A$ is defined as the angular
$p$-integral and the spatial integral of the coefficient
$\sigma^h_{-4}[A]$ in the homogeneous asymptotic expansion,
\begin{equation*}
\textrm{Res}(A):={\textstyle \frac1{(2\pi)^4}}
\int_{|p|=1}d\Omega_p\,\int_{\mathbbm R^4} \,d^4x\,
\sigma_{-4}^h[A](x,p),
\end{equation*}
whenever the integral exists. It is known that for compact spatial
manifolds this quantity determines completely the factor in front of
the $\log\Lambda$ term in the asymptotic expansion of
$\Tr_\Lambda(A)$. By abuse of notation, and motivated by the above
observation, in our calculations we will use the expression
Res$(\ldots)$ to mean the factor in front of the $\log\Lambda$ term in
the corresponding cut-off regularized trace.

\section{The case $M=\mathbbm R^4$}
\subsection{The logarithmically divergent part\label{logdiv}}

In this section we compute the logarithmically divergent part of the
bosonic effective action on $\mathbbm R^4$. We define the regularized
bosonic action as
\begin{equation}
  S_\Lambda(A) := \Tr_\Lambda
  \big({\textstyle \log(\frac{-\Box_A +
      m^2}{\Lambda_0^2}) - \log(\frac{-\Box_0 + m^2}{\Lambda_0^2})}\big).
\end{equation}
We use the following expression for the logarithm
\begin{equation}\label{DefLog}
\log (1+a) = \int_0^1 \frac{ds}{s} (1-(1+sa)^{-1})
\end{equation}
and recall the definition for the regularized trace of a
pseudo-differential operator
\begin{equation}
\Tr_\Lambda (a) := \int_{|p| \leq \Lambda}\frac{d^4
p}{(2\pi)^4} \int_{\mathbbm R^4} d^4x \,\mathrm{tr} \sigma[a] (x,p)
\end{equation}
to get
\begin{equation}
  \begin{split}
    \Tr_\Lambda
    &( {\textstyle \log (\frac{-\Box_A + m^2}{\Lambda_0^2})
      - \log (\frac{-\Box_0 + m^2}{\Lambda_0^2})}) \\
    &=
    -\int_{|p| \leq \Lambda} \frac{d^4p}{(2\pi)^4} \int_{\mathbbm R^4}
    d^4x \int_0^1 \frac{ds}{s} \TrN \big(\sigma[(I+s( {\textstyle
    \frac{-\Box_A +m^2}{\Lambda_0^2}} -I ))^{-1}] \\
    & \hspace*{16em}
    -
    \sigma[(I+s({\textstyle \frac{-\Box_0 + m^2}{\Lambda_0^2}}-
    I))^{-1}]
    \big)
  \end{split}
\end{equation}
As shown in Appendix \ref{App:clog}, the symbol of the resolvent of
$\Box_A$ has to satisfy the following recursion relation
\begin{equation*}
  \begin{split}
    \sigma[(c_1 I &+ c_2 \Box_A)^{-1}] (p,x) \\
    &=
    \textstyle
    \frac{1}{c_1I -
      c_2p^2} - \frac{c_2}{c_1I - c_2p^2} (\Box_A
    +2ip_\mu D^\mu_A) \sigma[(c_1 I + c_2 \Box_A)^{-1}] (p,x).
  \end{split}
\end{equation*}
Its formal solution is given by
\begin{equation*}
  \sigma[(c_1 I + c_2 \Box_A)^{-1}](x,p)
  =\big(c_1 I+c_2(-p^2+ \Box_A+2i\,p^\mu D_{A\mu})\big)^{-1}1
\end{equation*}
which can be understood as defining an asymptotic expansion, see the
appendix for details. In particular,
for our values of $c_1$ and $c_2$, we derive
\begin{equation*}
\begin{split}
  \sigma[(I + s[{\textstyle \frac{-\Box_A + m^2}{\Lambda_0^2}} - I])^{-1}]
  &\sim \sum_{n=0}^\infty \frac{ (s/\Lambda_0^2)^n}{(1-s+
    \frac{sm^2}{\Lambda_0^2} + \frac{s}{\Lambda_0^2} p^2)^{n+1}}
  (\Box_A +2ip_\mu D_A^\mu)^n 1.
\end{split}
\end{equation*}
Here and in all what follows, the $1$ on the r.h.s. means that the
operators $\Box_A$, $D_{A,\mu}$ should be applied to the $N$
dimensional constant vector.

Inserting this expansion into the integral and noting that the
second symbol just cancels the first term in the expansion we then
have
\begin{equation} \label{trlog}
  \begin{split}
    &\Tr_\Lambda\big({\textstyle \log(\frac{-\Box_A + m^2}{\Lambda_0^2}) -
      \log(\frac{-\Box_0 + m^2}{\Lambda_0^2})}\big)
    \\
    &= -\int_{|p| \leq
      \Lambda} \frac{d^4p}{(2\pi)^4} \sum_{n=1}^\infty
    \frac{1}{\Lambda_0^{2n}} \int_0^1 ds
     {\textstyle \frac{ s^{n-1}}{(1 + s(\frac{p^2 + m^2}{\Lambda_0^2}
         -1))^{n+1}}}
     \int_{\mathbbm R^4}d^4x\,\TrN(\Box_A + 2ip_\mu D_A^\mu)^n1.
  \end{split}
\end{equation}
In Appendix \ref{App:clog} we will expand explicitly the terms in
(\ref{trlog}) and pick out the logarithmically diverging ones.
Putting all the relevant terms together then gives
\begin{equation}
  \begin{split}
    &\mathrm{Res} ( \log (
    {\textstyle  \frac{-\Box_A + m^2}{\Lambda_0^2}) - \log
      (\frac{-\Box_0 + m^2}{\Lambda_0^2})})
    \\
    &= - \frac{1}{8\pi^2} m^2
    \int_{\mathbbm R^4} d^4x \,\TrN \Box_A - \frac{1}{16\pi^2}
    \int_{\mathbbm R^4} d^4x \,\TrN \Box_A^2 + \frac{1}{8\pi^2}
    m^2 \int_{\mathbbm R^4} d^4x \,\TrN \Box_A
    \\ & \quad + \frac{1}{12\pi^2} \int_{\mathbbm R^4} d^4x
    \TrN \Box_A^2 + \frac{1}{24\pi^2} \int_{\mathbbm R^4} d^4x
    \TrN D_A^\mu \Box_A D_{A\mu}
    \\ & \quad - \frac{1}{48\pi^2}
    \int_{\mathbbm R^4} d^4x \,\TrN (\Box_A^2 + D_A^\nu D_A^\mu
    D_{A\nu} D_{A\mu} + D_A^\mu \Box_A D_{A\mu})
    \\ &=
    \frac{1}{48\pi^2} ( \int_{\mathbbm R^4} d^4x \,\TrN D_A^\mu
    \Box_A D_{A\mu} - \int_{\mathbbm R^4} d^4x \,\TrN D_A^\nu
    D_A^\mu D_{A\nu} D_{A\mu})
  \end{split}
\end{equation}
A short calculation shows that the terms under the trace are equal to
$\frac{e^2}2 F^{\mu\nu}F_{\mu\nu}$, so we finally get the result
\begin{equation}
  \mathrm{Res}\bigg( \log (\frac{-\Box_A + m^2}{\Lambda_0^2}) - \log
  (\frac{-\Box_0 + m^2}{\Lambda_0^2})\bigg)
  = \frac{e^2}{96 \pi^2}
  \int_{\mathbbm R^4} d^4x \,\TrN F^{\mu\nu} F_{\mu\nu}
\end{equation}
which proves Proposition \ref{prp:comm}.

\subsection{Comparison with fermion calculations\label{Fermion}}

To incorporate fermions, we have to extend the Hilbert space. We take
$\mathcal H_{\textrm{fermion}}=L^2(\mathbbm R^4)\otimes \mathbbm
C^N_{\textrm{color}}\otimes \mathbbm C^4_{\textrm{spin}}$, where the
last factor carries a representation of the Dirac gamma matrixes
$\gamma^\mu$, $\mu=1,\cdots,4$.

We begin by computing the square of the Dirac operator $\Ds$.
First some definitions
\begin{equation*}
  \begin{split}
    D_{A\mu}&=\partial_\mu + ieA_\mu \\
    \dAslash&=\gamma^\mu(\partial_\mu + ieA_\mu)
\end{split}
\end{equation*}
A short calculation yields the well-known formula
\begin{equation*}
  \begin{split}
    \big({\dAslash}\big)^2
    &
    \textstyle
    =\mathbbm{1}_4 \Box_A + \frac12 \gamma^\mu \gamma^\nu
    [D_{A\mu},D_{A\nu}]\\
    &
    =\mathbbm{1}_4 \Box_A +i e \fslash.
  \end{split}
\end{equation*}
Here, $\mathbbm 1_4$ denotes the $4\times 4$ unit matrix and
\begin{equation*}
  \textstyle
  \fslash := \frac 14 \sigma^{\mu\nu} F_{\mu\nu}
  =\frac12 \gamma^\mu \gamma^\nu F_{\mu \nu}
\end{equation*}
for the matrices $\sigma^{\mu\nu}:=\frac12[\gamma^\mu,\gamma^\nu]$. 
We use the above identity to obtain 
\begin{equation*}
  (-i\dAslash+im)(-i\dAslash-im)=-\big(\dAslash\big)^2+m^2=
  -\mathbbm 1_4 \Box_A -ie\fslash+m^2.
\end{equation*}
Taking the logarithm on both sides, for the left hand side we arrive at
\begin{equation*}
  \log {\textstyle  (\frac{-i{\dAslash} +
      im}{\Lambda_0})(\frac{-i{\dAslash}-im}{\Lambda_0}})=
  \log({\textstyle  \frac{-i{\dAslash}+im}{\Lambda_0}}) +
  \log({\textstyle  \frac{-i{\dAslash}-im}{\Lambda_0}}),
\end{equation*}
while the right hand side gives
\begin{equation*}
  \begin{split}
    \log\big(
    -\mathbbm 1_4 \Box_A -ie\fslash+m^2\big)
    &=
    \textstyle
    \log\big(\frac{\mathbbm 1_4(-\Box_A+m^2)}{\Lambda_0^2}\big)
    \big(\mathbbm 1_4-\frac{ie}{-\Box_A+m^2} \fslash\big)
    \\&
    \textstyle
    =\log\big(\frac{\mathbbm 1_4 (-\Box_A+m^2)}{\Lambda_0^2}\big)
    +\log\big(\mathbbm 1_4-\frac{ie}{-\Box_A+m^2} \fslash\big)
    \\&\hspace*{10em}
    +\textrm{ commutator terms. }
  \end{split}
\end{equation*}
The extra commutator terms can be computed from the
Baker-Campbell-Haus\-dorff formula.  

It is known that on compact manifolds the Wodzicki residue vanishes
on commutators \cite{WodzRes}.  We therefore expect that from the
above expression, the commutator terms will not contribute to the
logarithmically divergent part of the regularized trace. In 
Appendix \ref{App:Lang} it is shown explicitly that this is indeed the
case. Rather than using integration-by-parts arguments, this is readily
seen from the fact that the $\mathbbm C_{\textrm{spin}}$-trace over
$\fslash$ gives zero. Also, the pertinent contributions from the first
two terms of the right hand side are calculated in the appendix.

Furthermore, from Langmann's results \cite{Lang01} we know that
$\Tr_\Lambda \log(\frac{-i{\dAslash}+im}{\Lambda_0})$ is
independent of the sign of $m$, so we have
\begin{equation*}
  \begin{split}
    \textstyle
    2 \Tr_\Lambda \log \big(\frac{-i{\dAslash}+im}{\Lambda_0}\big)
    &=\textstyle
    4 \Tr_\Lambda \log
    \big(\frac{-\Box_A+m^2}{\Lambda_0^2}\big)
    +\frac{e^2}{16 \pi^2} \log \Lambda \int_{\mathbbm R^4} d^4x
    \,\mathrm{tr} (\fslash)^2 \\
    &\hspace*{15em}+\textrm{ terms finite in $\Lambda$,}
  \end{split}
\end{equation*}
where the trace $\textrm{tr}$ runs over both the $\mathbbm
C^N_{\textrm{color}}$ and the $\mathbbm C^4_{\textrm{spin}}$ parts.
Performing the trace over the $\gamma$-matrices yields
\begin{equation*}
  \mathrm{tr} (\fslash)^2
  =-2 \TrN F^{\mu \nu}F_{\mu \nu}
\end{equation*}
The result is then
\begin{equation*}
  \begin{split}
    \Tr_\Lambda &\log
    ({\textstyle \frac{-i{\dAslash}+im}{\Lambda_0}})
    \textstyle
    =2\Tr_\Lambda \log{\textstyle (\frac{-\Box_A+m^2}{\Lambda_0^2}})
    -\frac{e^2}{16 \pi^2} \log \Lambda
    \int_{\mathbbm R^4} d^4x \,\TrN F^{\mu \nu} F_{\mu \nu} +\ldots
    \\[1ex]
    &=\textstyle
    \big(\frac{e^2}{48 \pi^2}\int_{\mathbbm R^4} d^4x
    \,
    \TrN F^{\mu \nu} F_{\mu \nu}-\frac{e^2}{16 \pi^2} \int_{\mathbbm R^4}
    d^4x \,\TrN F^{\mu \nu} F_{\mu \nu}
    \big)\log \Lambda
    +\ldots
    \\[1ex]
    &
    \textstyle
    =
    -\frac{e^2}{24 \pi^2} \log \Lambda \int_{\mathbbm
      R^4} d^4x \,\TrN F^{\mu \nu} F_{\mu \nu}
    + \textrm{ terms finite in $\Lambda$,}
  \end{split}
\end{equation*}
in agreement with \cite{Lang01}.

\subsection{Dependence on the regularization scheme\label{regdep}}

So far we have been looking at the cut-off regularized determinant
\begin{equation}
\textstyle
S_\Lambda(A)=\Tr_\Lambda\left(\log(\frac{-\Box_A+m^2}{\Lambda_0^2})
-\log(\frac{-\Box_0+m^2}{\Lambda_0^2})\right).
\end{equation}
As the cut-off in this regularization is taken with respect to the
reference operator $\Box_0$, the above expression is not manifestly
gauge invariant. It is thus surprising that the coefficient
$c_{\textrm{log}}(A)$ turns out to be gauge invariant.

One could use the spectral projection with respect to $\Box_A$
instead, but again the resulting expression would fail to be
manifestly gauge invariant now because of the reference term
$\log(\frac{-\Box_0+m^2}{\Lambda_0^2})$. The latter had to be
included to make the calculations tractable by the methods of
classical \psdo s.

Of course, there are gauge invariant regularization schemes such as
heat kernel regularization \cite{Vass-Manual} readily
available. However, cut-off regularized traces seem to be closer to
physical intuition.

An acceptable, manifestly gauge invariant expression would be
\begin{equation}\label{det-gaugeinv}
  \textstyle
  \tilde S_\Lambda(A)
  :=
  \Tr_\Lambda^{\Box_A} \log(\frac{-\Box_A+m^2}{\Lambda^2_0})-
  \Tr_\Lambda\log(\frac{-\Box_0+m^2}{\Lambda^2_0}),
\end{equation}
where $\Tr_\Lambda^{\Box_A}\log(\frac{-\Box_A+m^2}{\Lambda^2_0}):=\Tr
\left\{P_\Lambda(\Box_A)\log(\frac{-\Box_A+m^2}{\Lambda^2_0})\right\}$
is defined using the spectral projections\footnote{In this section,
$\theta$ denotes the Heaviside step function that is zero for negative
arguments and equal to $1$ otherwise.}
$P_\Lambda(\Box_A):=\theta(\Lambda^2-\Box_A)$ of $\Box_A$. It turns
out that $\tilde S_\Lambda(A)$ 
asymptotic expansion as
\begin{equation}
  \tilde S_\Lambda(A)
  =
  \tilde c_2(A)\Lambda^2+\tilde c_1(A)\Lambda+
  \tilde c_{\textrm{log}}(A)\log\Lambda+\ldots
\end{equation}
The dots indicate terms that are finite in the large $\Lambda$ limit.\\
In this section we want to compare the coefficient $\tilde
c_{\textrm{log}}(A)$ of the logarithmically divergent part in the
above expression to the coefficient $c_{\textrm{log}}(A)$ computed
earlier.

A short calculation reveals how to proceed:
\begin{equation}\label{telescope}
  \begin{split}
    \Tr^{\Box_A}_\Lambda &\log \frac{-\Box_A+m^2}{\Lambda_0^2}
    -\Tr^{\Box_0}_\Lambda \log\frac{-\Box_0+m^2}{\Lambda^2_0}
    \\&
    =\Tr^{\Box_0}_\Lambda \left(\log \frac{-\Box_A+m^2}{\Lambda_0^2}
    -\log \frac{-\Box_0+m^2}{\Lambda_0^2}\right)\\
    &\quad+\left(\Tr^{\Box_A}_\Lambda-\Tr^{\Box_0}_\Lambda\right)
    \left(\log \frac{-\Box_A+m^2}{\Lambda_0^2}
    -\log \frac{-\Box_0+m^2}{\Lambda_0^2}\right)\\
    &\quad+\left(\Tr^{\Box_A}_\Lambda-\Tr^{\Box_0}_\Lambda\right)\log
    \frac{-\Box_0+m^2}{\Lambda_0^2}.
  \end{split}
\end{equation}
Obviously, the coefficient $\tilde c_{\textrm{log}}(A)$ receives
contributions from three different terms, only the first of which is
given by $c_{\textrm{log}}(A)$. From the calculation of the pertinent
part in the third term, it will be apparent that the second one in
fact does not contribute to $\tilde c_{\textrm{log}}(A)$. For the
computation of the third term in (\ref{det-gaugeinv}), however, we
have to introduce an additional regulator that deals with the
non-compactness of $\mathbbm R^4$. Let $f$ be a smooth, compactly supported
function on $\mathbbm R^4$, interpreted as a multiplication operator on
$\mathcal H$. Then
\begin{equation}
  \begin{split}
    \Tr \big\{
    f\,\theta(\Lambda^2+\Box_A)\,
    &{\textstyle \log\frac{-\Box_0+m^2}{\Lambda^2_0}}\big\}
    \\
    &=\int_{\mathbbm R^4}\frac{d^4p}{(2\pi)^4}\int_{\mathbbm R^4}
    \,f(x)\,\sigma[\theta(\Lambda^2+\Box_A)]\,
    \sigma[{\textstyle\log\frac{-\Box_0+m^2}{\Lambda^2_0}}]+\ldots
  \end{split}
\end{equation}
The dots indicate contributions from the star product of symbols
that are uniformly bounded in $\Lambda$. Use has been made of the
fact that $f$ and $\sigma[\log\frac{-\Box_0+m^2}{\Lambda^2_0}]$ are
independent of $p$ and $x$, respectively. 

As a next step, we need to derive an asymptotic expansion for the
symbol of the theta function. We start with the following sum
expression for a smooth approximation of the Heaviside $\theta$
function \cite[p. 248 {\sl et seq.}]{FetterWalecka}
\begin{equation}\label{Heaviside}
  \theta_\epsilon (x) = \frac1\epsilon
  \sum_{r=-\infty}^\infty \frac{e^{i\omega_r0^+}}{x+i\,\omega_r}
  =\frac{e^{-x0^+}}{e^{-x\epsilon}+1},\quad
  \omega_r=(2r+1)\pi /\epsilon,
\end{equation}
for $\epsilon> 0$. The step function is regained in the limit
$\epsilon\rightarrow \infty$. Using this equation, we derive an asymptotic
expansion for the symbol of $\theta(\Lambda^2+\Box_A)$ as (for details
we refer to the Appendix \ref{App:regdep})
\begin{equation*}
  \begin{split}
    \sigma [\theta_\epsilon (\Lambda^2 + \Box_A)] &= \frac{1}{2 \pi i} \int dz
    e^{iz\epsilon} \sigma \biggl[
      {\textstyle\frac{1}{z-i(\Lambda^2 + \Box_A)}}\biggr]\\
    &\sim\sum_{n=0}^{\infty}
    \frac{1}{n!} \delta^{(n-1)}_\epsilon(\Lambda^2 - p^2)(\Box_A+2ip^\mu
    D_{A\mu})1.
  \end{split}
\end{equation*}
As before, for mnemonic purposes, this asymptotic series can be
summarized as
\begin{equation}
\sigma\left[\theta(\Lambda^2+\Box_A)\right](x,p)=
\theta(\Lambda^2-p^2+\Box_A+2i \,p^\mu \,D_{A\mu})1,
\end{equation}
where the $x$ dependence originates from the external fields $A$.
\\
Combining our results, we find
\begin{equation*}
  \begin{split}
    \Tr&\left\{ f\,\theta(\Lambda^2+\Box_A)\,
       {\textstyle \log (\frac{- \Box_0 +m^2}{\Lambda_0^2})}\right\}
       -
       \Tr\left\{
       f\,\theta(\Lambda^2+\Box_0)\,
       {\textstyle \log (\frac{- \Box_0
	   +m^2}{\Lambda_0^2})
       }
       \right\} \\ 
       &= \sum_{n=1}^{\infty}\frac{1}{n!} \int_{\mathbbm R^4}
       \frac{d^4p}{(2\pi)^4} 
       \int_{\mathbbm R^4} d^4x\,f(x)\,
       \delta^{(n-1)}_\epsilon 
       (\Lambda^2-p^2)
       \,{\textstyle \log(\frac{p^2+m^2}{\Lambda_0^2})} \times\\
       &\hspace*{17em}
       \times\TrN \left\{(\Box_A + 2ip^\mu D_{A \mu})^n1\right\}
  \end{split}
\end{equation*}
Obviously, we can now drop the regulator $f$.

For large $\Lambda$, the $\delta_\epsilon$-functions cancel the
radial p-integration. Therefore, the only contributions to the
logarithmically divergent part in the above expression can
originate from terms where the derivatives of the
$\delta_\epsilon$-functions exclusively hit the trace under the
integral of the measure $d^4p$ but not the factor
$\log\frac{-\Box_0+m^2}{\Lambda^2_0}$. This is only possible as long
as $2(n-1)\leq 3+n$ (the derivatives of $\delta_\epsilon$ count
twice because of the $p^2$ in the argument, and the $3$ on the
r.h.s. comes from the measure $d^4p$) and hence $n\leq 5$.
Moreover, since the angular $p$-integration over an odd number of
factors $p^\mu$ gives always zero, the $n=5$ term  cannot
contribute either.

As shown in the Appendix, we can now expand the powers
$(\Box_A+2\,i\,p^\mu\,D_{A\mu})^n1$ for $n\leq 4$, perform the
angular $p$-integrations, substitute $p^2\rightarrow u$ and use
partial integration to get rid of the derivatives of the
$\delta_\epsilon$-functions. We arrive at
\begin{equation*}
  \begin{split}
    &\Tr_\Lambda^{\Box_A} \log (\frac{- \Box_0 +
      m^2}{\Lambda_0^2}) - \Tr_\Lambda^{\Box_0} \log (\frac{-\Box_0
      + m^2}{\Lambda_0^2}) \\
    &=\frac{1}{16\pi^2} \int_1^{\infty} du \log
    ({\textstyle \frac{u + m^2}{\Lambda_0^2}})
    \delta_\epsilon(\Lambda^2-u) u (1-1)
    \int_{\mathbbm R^4}d^4x \,\TrN\Box_A1 \\
    & \quad +
    \frac{1}{16\pi^2} \int_1^{\infty} du
    \log ({\textstyle \frac{u + m^2}{\Lambda_0^2}})
    \delta_\epsilon(\Lambda^2-u)
    (\frac12-\frac23+\frac16) \int_{\mathbbm R^4}d^4x
    \TrN\Box_A^21 \\
    & \quad + \frac{1}{16\pi^2} \int_1^{\infty}
    du \log ({\textstyle \frac{u + m^2}{\Lambda_0^2}})
    \delta_\epsilon(\Lambda^2-u)
    (-\frac13+\frac16) \int_{\mathbbm R^4}d^4x \,\TrN D_A^\mu
    \Box_A D_{A\mu}1 \\
    & \quad + \frac{1}{16\pi^2} \int_1^{\infty} du
    \log ({\textstyle \frac{u + m^2}{\Lambda_0^2}}) \frac16
    \delta_\epsilon(\Lambda^2-u)
    \int_{\mathbbm R^4}d^4x
    \TrN D_A^\mu D_A^\nu D_{A\mu} D_{A\nu}1
    +\ldots\\
    &=-\frac{1}{96\pi^2} \log
    ({\textstyle \frac{\Lambda^2 + m^2}{\Lambda_0^2}}) \int_{\mathbbm R^4}d^4x
    \TrN D_A^\mu \Box_A D_{A\mu}1  \\
    & \quad +
    \frac{1}{96\pi^2} \log ({\textstyle \frac{\Lambda^2 + m^2}{\Lambda_0^2}})
    \int_{\mathbbm R^4}d^4x \,\TrN D_A^\mu D_A^\nu D_{A\mu}
    D_{A\nu}1
    +\ldots\\
    &=-\frac12\frac{1}{96\pi^2}
    \log({\textstyle \frac{\Lambda^2+m^2}{\Lambda^2_0}})
    \int_{\mathbbm R^4}\,d^4x \,\TrN F^{\mu\nu} F_{\mu\nu}+\ldots,
  \end{split}
\end{equation*}
where the dots indicate finite or polynomially divergent
contributions.

\noindent Finally, we will turn back to the second term in
(\ref{telescope}). The difference as compared to the previous
calculation is that now the symbol of the operator under the
traces has an asymptotic expansion that is a power series in
$\frac1p$. Therefore, in contrast to the above, no logarithmically
divergent term will occur in a large $\Lambda$ expansion.
\\
Combining these results with our previous expression for
$c_{\textrm{log}}(A)$, we find
\begin{equation}
\tilde c_{\textrm{log}}(A)=
\frac12\frac{1}{96\pi^2}
\int_{\mathbbm R^4}\,d^4x \,\TrN  F^{\mu\nu} F_{\mu\nu} =\frac12
c_{\textrm{log}}(A).
\end{equation}

\subsection{Comparison with heat kernel regularization\label{heatkernel}}

In this section we want to compare our results with previous ones in
the literature \cite{Vass03,Moyal} obtained by heat kernel
techniques. 
For a given differential operator $D$, we consider the
trace of the heat operator for $D$, 
\begin{equation*}
K(t,f,D)=\Tr (fe^{-tD}),
\end{equation*}
where the auxiliary smooth function $f(x)$ is introduced to make
spatial integrals converge on $\mathbbm R^n$. We write the
effective action for $D$ as
\begin{equation*}
S=-\int_0^\infty \frac{dt}{t} K(t,f,D).
\end{equation*}
Here the formula $\log \det (D)= \Tr \log (D)$ has been used again
together with the following formal expression for the
logarithm
\begin{equation*}
\log \lambda =-\int_0^\infty \frac{dt}{t} e^{-t \lambda}
\end{equation*}
which holds up to an (infinite) integration constant.\\
There is an asymptotic expansion for the heat trace as $t
\rightarrow 0$ given by
\begin{equation*}
\Tr(fe^{-tD}) \sim \sum_{k \geq 0} t^{(k-n)/2}a_k(f,D).
\end{equation*}
Next we define the $\zeta$-function for $D$ as follows
\begin{equation*}
\zeta (s,f,D)=\Tr(fD^{-s}).
\end{equation*}
Writing the $\zeta$-function in terms of the heat trace as
\begin{equation*}
\zeta (s,f,D)=\frac{1}{\Gamma(s)} \int_0^\infty dtt^{s-1}
K(t,f,D).
\end{equation*}
we see that $\Gamma (s) \zeta (s,f,D)$ has simple poles at the
points $s=(n-k)/2$ and the complex residue at $s=(n-k)/2$ is given by
\begin{equation}\label{calc-a}
\mathrm{Res}_{s=(n-k)/2}(\Gamma (s) \zeta (s,f,D))=a_k(f,D)
\end{equation}
From the asymptotic expansion of the heat trace and the integral
formula for the effective action $S$ we see that the
logarithmically divergent part is given when $k=n$ so we are
interested in computing the coefficient
$a_n(f,\textstyle{\frac{-\Box_A+m^2}{\Lambda_0^2}}).$ In our case
$n=4$.

The first task is to compute the $\zeta$-function for the operator
$(-\Box_A+m^2)/\Lambda_0^2$. From the definition of the
$\zeta$-function we have
\begin{equation*}
  \zeta (s,f,{\textstyle \frac{-\Box_A +m^2}{\Lambda_0^2}})
  =\int_{\mathbbm R^4}
  \frac{d^4p}{(2 \pi)^4} \int_{\mathbbm R^4} d^4x \TrN
  \sigma[f] \ast
  \sigma[({\textstyle \frac{-\Box_A+m^2}{\Lambda_0^2}})^{-s}](x,p).
\end{equation*}
We next use the expansion
\begin{equation*}
  (a+x)^{-s}=\sum_{r=0}^\infty (-1)^r \frac{\Gamma (s+r)}{r! \Gamma
    (s)}a^{-(r+s)}x^r
\end{equation*}
to write the symbol of $(-\Box_A+m^2)/\Lambda_0^2$ as
\begin{equation*}
  \begin{split}
    \sigma \left[ \left(
      {\textstyle\frac{-\Box_A+m^2}{\Lambda_0^2}}\right)^{-s} 
      \right]
    &
    =\bigg(\frac{p^2+m^2-\Box_A-2ip^{\mu}D_{A\mu}}{\Lambda_0^2}\bigg)^{-s}
    \\
    &
    \sim\sum_{r=0}^\infty (-1)^r \frac{\Gamma (s+r)}{r!\Gamma
      (s)}\Lambda_0^{2s} \frac{1}{p^{2(s+r)}}(m^2-\Box_A-2ip^\mu
    D_{A\mu})^r1.
\end{split}
\end{equation*}
Splitting the integration in the $\zeta$-function into two parts we
then have
\begin{equation} \label{zetaint}
  \begin{split}
    \zeta
    &
    (s,f,{\textstyle \frac{-\Box_A +m^2}{\Lambda_0^2}})
    =
    \int_{|p|\leq 1} \frac{d^4p}{(2 \pi)^4}
    \int_{\mathbbm R^4} d^4x f(x) \TrN  \sigma
    \bigl[\big({\textstyle \frac{-\Box_A+m^2}{\Lambda_0^2}}\big)^{-s}
      \bigr](x,p)+\\
    &
    +\sum_{r=0}^\infty
    \frac{(-1)^r \Gamma (s+r)}{r! \Gamma (s)}\Lambda_0^{2s}
    \int_1^\infty |p|^3d|p| \frac{1}{p^{2(r+s)}}\times\\
    &\hspace*{6em}\times
    \int_{S^3} \frac{d \Omega_p}{(2\pi)^4}
    \int_{\mathbbm R^4}d^4x f(x)\TrN (m^2-\Box_A-2ip^\mu D_{A\mu})^r1.
  \end{split}
\end{equation}
Using the fact that under the angular integration odd powers of
$p$ give zero we can write
\begin{equation*}
  \int_{S^3} \frac{d \Omega_{\xi}}{(2\pi)^4}\int_{\mathbbm R^4}d^4x\,
  f(x)\TrN (m^2-\Box_A-2ip^\mu
  D_{A\mu})^r1=\sum_{k=0}^{[r/2]}(-2i)^{2k}p^{2k}d(f,r,2k)
\end{equation*}
for some functions $d(f,r,2k)$ of $f$, $\Box_A$ and $D_{A\mu}$
determined from the expansion of $(m^2-\Box_A-2ip^\mu D_{A\mu})^r$. In
particular, we have
\begin{equation*}
  d(f,0,0)=\frac{1}{8\pi^2}\int_{\mathbbm R^4}\,d^4x\,f(x).
\end{equation*}
We then find
\begin{equation*}
  \begin{split}
    &\zeta (s,f,\textstyle \frac{-\Box_A +m^2}{\Lambda_0^2})=\chi
    (s)+\\
    &+\sum_{r=0}^\infty \frac{(-1)^r \Lambda_0^{2s}\Gamma
      (s+r)}{r! \Gamma (s)}\int_1^\infty
    |p|^3d|p|\frac{1}{p^{2(r+s)}}
    \sum_{t=0}^{[r/2]}(-2i)^{2t}p^{2t}d(f,r,2t).
  \end{split}
\end{equation*}
where $\chi (s)$ denotes the first integral in the rhs of
(\ref{zetaint}), a holomorphic function in $s$. We can evaluate
explicitly the $p$-integral in the above expression to obtain the
following formula for the $\zeta$-function.
\begin{equation}\label{zeta-d}
  {\textstyle
  \zeta (s,f,\frac{-\Box_A +m^2}{\Lambda_0^2})}
  =\chi(s)
  +{\textstyle\frac{\Lambda_0^{2s}}{\Gamma (s)}}
  \sum_{r=0}^\infty
  \sum_{t=0}^{[r/2]}{\textstyle \frac12 \frac{(-1)^{r+t}4^t \Gamma
    (s+r)}{r!}\frac{1}{s-(2-r+t)}} d(f,r,2t).
\end{equation}
There are two parts of the $\zeta$-function contributing to the residue
at $s=0$; the gamma function $\Gamma(s+r)$ and the poles of
$\frac{1}{s-(2-r+t)}$. The first one gives a contribution for $r=0$
and the latter one when $r=2+t$. From the summation we see that $t\leq
r/2$ so it follows that only the terms with $t\leq 2$ contribute to
the residue.
\begin{equation*}
  \begin{split}
    a_4(f,{\textstyle \frac{-\Box_A+m^2}{\Lambda_0^2}})
    =\mathrm{Res}_{s=0}\Gamma
    (s) \zeta (s,f,{\textstyle \frac{-\Box_A+m^2}{\Lambda_0^2}})
    &=\chi(0) -\frac14
    d(f,0,0) \\
    &\quad+
    \sum_{t=0}^2 \frac{4^t \Gamma (2+t)}{(2+t)!}\frac12
    d(f,t+2,2t)
\end{split}
\end{equation*}
where $\chi(0)$ is given by
\begin{equation*}
  \chi(0)=\frac{1}{(2 \pi)^4} \int_{|p|\leq 1}d^4p \int_{\mathbbm
    R^4} d^4x\, f(x)= \int_0^1 |p|^3 d|p| d(f,0,0)=\frac14 d(f,0,0).
\end{equation*}
We have thus obtained the following expression for
$a_4(f,{\textstyle \frac{-\Box_A+m^2}{\Lambda_0^2}})$
\begin{equation*}
  a_4(f,{\textstyle \frac{-\Box_A+m^2}{\Lambda_0^2}})
  =\sum_{t=0}^2 \frac12 \frac{4^t \Gamma (2+t)}{(2+t)!}
d(f,t+2,2t).
\end{equation*}

We now compute directly the logarithmically divergent part of
$S_\Lambda(A)$. For this we need the following formula
\begin{equation*}
  \begin{split}
    \sigma \biggl[ \log (\frac{-\Box_A+m^2}{\Lambda_0^2}) -
      & \log (\frac{-\Box+m^2}{\Lambda_0^2})\biggr]
    \\
    &=\sum_{r=1}^\infty \frac{(-1)^{r+1}}{r}
    \frac{1}{p^{2r}}[(m^2-\Box_A-2ip^\mu D_{A\mu})^r-m^{2r}]
  \end{split}
\end{equation*}
Note that for this asymptotic expansion, we have divided the
recursion formula (\ref{resolvent}) differently.
\\
We split the $p$-integration in the trace into two parts to get
\begin{equation*}
  \begin{split}
    \Tr_\Lambda \log\big(f&(\frac{-\Box_A+m^2}{\Lambda_0^2}
    -\frac{-\Box_0+m^2}{\Lambda_0^2})\big)
    \\
    &=\textrm{ finite terms in $\Lambda$ }+
    \\
    &\quad
    + \sum_{r=1}^\infty \sum_{t=0}^{[r/2]} \frac{(-1)^{r+1}}{r}
    \int_1^\Lambda |p|^3 d|p| \frac{1}{p^{2(r-t)}}
    (-1)^t4^td(f,r,2t)\\
    &\quad
    -\sum_{r=1}^{\infty} \frac{(-1)^{r+1}}{r}
    m^{2r}\int_1^\Lambda |p|^3d|p| \frac{1}{p^{2r}}d(f,0,0).
  \end{split}
\end{equation*}
The logarithmically divergent part is then given by
\begin{equation*}
  \mathrm{c}_{\log}(A)
  =-\sum_{t=0}^2
  4^t\frac{1}{t+2}d(f,t+2,2t)+\frac12m^4d(f,0,0)
\end{equation*}
so we finally have the result
\begin{equation*}
  -\frac12 \mathrm{c}_{\log}(A)+\frac14m^4d(f,0,0)
  =
  a_4(f,{\textstyle \frac{-\Box_A+m^2}{\Lambda_0^2}}).
\end{equation*}
{\bf Remarks.}
1) The coefficients $d(f,r,2k)$ defined below Eg.~(\ref{zetaint}) are
given by spatial integrals over the $gl(N)$-trace of 
certain polyomials in the external fields and their derivatives. They
can be easily computed by expanding the power on the left hand side of
the defining formula, using the well-known expressions for the
angular $p$-integration of polynomials in $p^\mu$.\\
2) Note that the argument relating $a_{-4}$ and $c_{\textrm{log}}$ did
not use the specific form of the coefficients $d(f,r,2k)$. Therefore,
it can be extended to a larger class of operators.\\
3) Combining Eqs (\ref{calc-a}) and (\ref{zeta-d}), we have a formula
for the calculation of the coefficients $a_k$ at hand. In particular,
evaluating the function $\chi(s)$ for negative integer $s$ amounts to
the computation of the symbol of $(-\Box_A+m^2)^l$ for positive
integer powers of $l$. The latter can be obtained from the formula
\begin{equation*}
\sigma\big[(-\Box_A+m^2)^{l+1}\big]
  =(p^2+m^2-\Box_A-2\,i\,p_\mu\,D^\mu_A)
\sigma\big[(-\Box_A+m^2)^{l}\big]
\end{equation*}
and the symbol of $-\Box_A+m^2$.

\section{Generalization to the Moyal plane}

\subsection{The Moyal plane $\mathbbm R^4_\theta$: generalities}

In this section, we want to replace the manifold $\mathbbm R^4$ by the
four-dimensional (4D) Moyal plane $\mathbbm R^4_\theta$, an example of
a noncommutative manifold.

For the definition of the latter, one has to specify (among other
things; see \cite{Connes} for the general theory,
\cite{SpectralTriples} for the treatment of the Moyal plane in this
context) a (noncommutative) associative algebra $\mathcal A$, the
elements of which generalize the notion of (smooth) functions on an
ordinary manifold. In the case of the 4D Moyal plane, the algebra
$\mathcal A$ is taken to include the rapidly decaying Schwartz test
functions on $\mathbbm R^4$, while the product of two such elements
$f$, $g$ is given by the integral formula
\begin{equation}\label{MoyalProduct}
(f\star g)(x)=\frac1{(2\pi)^4}\int_{\mathbbm R^4}\int_{\mathbbm R^4}
\, d^4y\,d^4\xi\,e^{i\xi(x-y)}\,f(x-{\textstyle
  \frac12}\Theta\xi)\,g(y),
\end{equation}
where $\Theta$ is a $4\times 4$ matrix defined by
\begin{equation}
  \Theta=\theta
  \left(\begin{array}{cc}0&\mathbbm 1_2\\
    -\mathbbm 1_2&0
        \end{array}
  \right).
\end{equation}
for the real parameter $\theta$.

The elements of $\mathcal A$ act on the Hilbert space $L^2(\mathbbm
R^4)$ by left $\star$-multiplication (see \cite{EGBV} for an extension
of the above formula to distributions). For an element $f\in\mathcal
A$, we will write the corresponding operator on $L^2(\mathbbm R^4)$ as
$f\star$.
From the integral formula (\ref{MoyalProduct}) of $\star$, we can see
that $f\star$ is a \psdo\ with the symbol
\begin{equation}
\sigma[f\star](x,p)=f(x-{\textstyle\frac12}\Theta p).
\end{equation}
Note that the asymptotic behavior of $f$ is transferred to the $p$
dependence of the symbol of $f\star$. In particular, for rapidly
decaying $f$, $f\star$ is infinitely smoothing
\cite{SpectralTriples}.

A natural class of functions suitable for the Moyal product is the set
$\mathcal P$ of infinitely differentiable functions $f$ on $\mathbbm R^4$
such that, for a real number $s$ and for every multi-index $\alpha$,
\begin{equation}\label{P-order}
|(\partial^\alpha_xf)(x)|\leq C_\alpha (1+x^2)^{\frac{s-|\alpha|}2}.
\end{equation}
$s$ is called the {\em order} of $f$. For $f$, $g$ $\in$ $\mathcal P$ and
of order $s_1$, $s_2$, respectively, $f\star g$ is again in $\mathcal P$ and
of order $s_1+s_2$ (\cite[sect. 7]{GLS}).

\subsection{Calculation of the logarithmically divergent part}

With the commutative product of functions on $\mathbbm R^4$ replaced
by the Moyal product $\star$, Eg.~(\ref{MoyalProduct}), we are led to
study the following variant of the Klein-Gordon operator
\begin{equation*}
\Box_A^\theta \psi =
\partial^\mu
\partial_\mu \psi + ie(\partial^\mu A_\mu) \star \psi + 2ie A^\mu
\star
\partial_\mu \psi - e^2 A^\mu \star A_\mu \star \psi
\end{equation*}
for any rapidly decaying smooth function $\psi$ in the Hilbert space
$\mathcal H=L^2(\mathbbm R^4)\otimes \mathbbm
C^N_{\textrm{color}}$. Here, the matrix valued Yang-Mills fields
$A_\mu$ are taken to be in the set $\mathcal P$ above with order strictly
smaller than $-4$, i.e. to satisfy (\ref{P-order}) with $s<-4$.

\noindent
We will also need the operator $D^\theta_{A\mu}$, defined by
\begin{equation*}
D_{A\mu}^\theta \psi =  \partial_\mu \psi + ie A_\mu \star \psi,\quad
\psi\in\mathcal S(\mathbbm R^4).
\end{equation*}
In analogy with the first section, we consider the cut-off regularized
determinant of $\frac{-\Box^\theta_A+m^2}{\Lambda^2_0}$,
\begin{equation*}
\textstyle
S^\theta_\Lambda(A):=\Tr_\Lambda
\left\{
\log\frac{-\Box_A^\theta+m^2}{\Lambda^2_0}
-\log\frac{-\Box_0+m^2}{\Lambda^2_0}
\right\}.
\end{equation*}
As before, the trace will be computed from the symbol of
$\log\frac{-\Box_A^\theta+m^2}{\Lambda^2_0}$. For the latter, we will
need an expression for the symbol of the resolvent of
$\Box_A^\theta$. Again, this will be obtained via a recursion
relation.

As explained in the Appendix, we find for $c_1$, $c_2$ $\in$ $\mathbbm
C$,
$c_1\cdot c_2<0$ or $c_2=0$,
\begin{equation*}
  \begin{split}
    \sigma[&(c_1+c_2\Box_A^\theta)^{-1}](x,p)
    \\
    &\textstyle=\frac1{c_1-c_2p^2}
    -
    \frac {c_2}{c_1-c_2p^2}
    \left(\Box^\theta_{A(\cdot-\frac12\Theta p)}
    +2 i\,p^\mu\,D^\theta_{A(\cdot-\frac12\Theta p)}\right)
    \sigma[(c_1+c_2\Box_A^\theta)^{-1}](x,p),
  \end{split}
\end{equation*}
where $A(\cdot-\frac12\Theta p)$ is a short hand notation for the
external fields $A_\mu$ shifted by $-\frac12\Theta p$ in their
argument,
$A(\cdot-\frac12\Theta p)(x)=A(x-\frac12\Theta p)$.
In the derivation of the recursion relation, we have used the identity
\cite{Kammerer}
\begin{equation*}
  \textstyle
  e^{i p(x-y)}f(x)=[f(\cdot+\frac12\theta)\star e^{i p(\cdot-y)}](x)
\end{equation*}
and associativity of the Moyal product.

From the recursion relation, one readily obtains the formal expression
\begin{equation*}
  \sigma[(c_1+c_2\Box^\theta_A)^{-1}](x,p)
  \sim
  \sum_{n=0}^\infty
      {\textstyle \frac{(-1)^n}{(c_1-c_2p^2)^{n+1}}}
      \left(\Box^\theta_{A(\cdot-\frac12\Theta p)}
      +2 i\,p^\mu\,D^\theta_{A(\cdot-\frac12\Theta p)}\right)^n1.
\end{equation*}
A thorough investigation reveals, however, that an interpretation of
this equation as an asymptotic expansion in $p$ would be misleading:
The $p$ dependence through the arguments of the external fields
$A_\mu(x-\frac12\Theta p)$ is superficial in that it goes away under
the spatial integral.  Therefore, one has to develop different tools
to tackle the situation. As shown in the appendix, the operator $R_N$
defined by the sum of the first $N$ terms in the above series, for $N$
sufficiently large, differs from the operator
$(c_1+c_2\Box^\theta_A)^{-1}$ by a trace class operator only. Hence,
for the singular behavior of the cut-off regularized trace, it
suffices to consider this operator $R_N$.

Inserting the expression for the symbol of $R_N$ into the
integral formula for the logarithm, Eg.~(\ref{DefLog}), we find
\begin{equation*}
  \begin{split}
    S_\Lambda^\theta(A)&=
    -\sum_{n=1}^N
    \int_{|p|\leq \Lambda}{\textstyle \frac{d^4p}{(2\pi)^4}}
    \int_0^1{\textstyle \frac{ds}s}
    {\textstyle
      \frac{s^{n-1}}{(1+s(\frac{p^2+m^2}{\Lambda^2_0}-1))^{n+1}}}
    \times\\
    &\hspace*{7em}\times\int_{\mathbbm R^4}
    d^4x\,
    \TrN(\Box^\theta_{A(\cdot-\frac12\Theta p)}+2 i
    p^\mu\,D^\theta_{A(\cdot-\frac12\Theta p),\mu})^n1\\[1ex]
    &\hspace*{20em}\textrm{+terms finite in $\Lambda$.}
  \end{split}
\end{equation*}
Now, for every term in the sum, we can shift the $x$-integration by
$-\frac12\Theta p$. After this substitution the contribution to the
$\Lambda$-behavior is apparent: It is only the first four terms that
can contribute to $c_{\textrm{log}}^\theta(A)$. Moreover, the
resulting expression differs from the corresponding for
$S_\Lambda(A)$, Eg.~(\ref{trlog}), solely in the appearance of the
product $\star$ in place of the commutative product. As the
replacement of the latter by the Moyal product does not affect the
asymptotic behavior in the variable $p$, we conclude
\begin{equation}
  c_{\textrm{log}}^{\theta}(A)
  ={\textstyle \frac {e^2}{96\pi^2}}
  \int_{\mathbbm R^4}\,d^4x\,
  \TrN F^{\theta,\mu\nu}\star F^\theta_{\mu\nu},
\end{equation}
where $F^\theta_{\mu\nu}$ is defined by
\begin{equation*}
  F^\theta_{\mu\nu}
  =-{\textstyle ie}[D^\theta_{A\mu},D^\theta_{A\nu}].
\end{equation*}
This proves the claim of Proposition \ref{prp:star}.

\section{Conclusion}

In the first part of our paper, we considered the regularized
determinant of the Klein-Gordon operator $\Box_A$ with minimal
coupling on $\mathbbm R^4$. For the regularization, we restricted the
Hilbert space trace to run over states of momentum below some cut-off
$\Lambda$.
 
Although similar results have been obtained before, we chose to
present here an approach that consistently uses pseudodifferential
operator methods to prepare the ground for calculations on a
particular noncommutative manifold. 

A useful formula for the
calculations with symbols of pseudodifferential operators (\psdo) is
given by
\begin{equation}\label{concl:c-symbol}
  \sigma\big[f(\Box_A)\big](x,p)=f(-p^2+\Box_A+2i\,p^\mu\,D_{A\mu})1
\end{equation}
for any function $f$ of the Klein-Gordon operator $\Box_A$. This
formula originates from a recursion relation for the symbol
$\sigma\big[f(\Box_A)\big]$. It is to be understood as defining an
asymptotic expansion of the symbol for large $p$.

Using this asymptotic expansion we could indeed confirm that the
cut-off regularized trace does have an asymptotic expansion in the
cut-off $\Lambda$ as in Eg.~(\ref{S-asymptext}).  Although our
approach did not use a manifestly gauge invariant regularization, the
term scaling like $\log\Lambda$ in the regularized trace of the
logarithm of the massive Klein-Gordon operator was found to be gauge
invariant. However, the numerical coefficient in front of this
expression differs from that obtained via manifestly gauge invariant
methods \cite{deWitt,Moyal} by a factor of minus two, see Eqs
(\ref{res:regdep}) and (\ref{heat-kernel-result}). This difference can
be understood through a comparison of our approach to heat kernel
regularization. It turns out that this argument does not rely on the
particular structure of the operator $\Box_A$, cf. the use of the
functions $d(f,r,2t)$ in Section \ref{heatkernel}, so we expect it to
hold even for more general operators as well. It would be interesting
to understand this feature in more detail. Also, we propose a gauge
invariant version of the cut-off regularization,
Eg.~(\ref{ginvdet-defn}),which reproduces the result of
\cite{deWitt,Moyal}.

In the second and main part of our work, we considered the generalized
Klein-Gordon operator for minimally coupled bosons on the four
dimensional Moyal plane, a particular example for a noncommutative
geometry. The difference to the previous case is that now the external
Yang-Mills fields act on wave functions by the noncommutative Moyal
multiplication. This leads in a natural way to the generalized
Klein-Gordon operator $\Box_A^{\theta}$. As it turns out, the
machinery of \psdo s is still applicable, with (\ref{concl:c-symbol})
generalizing to
\begin{equation}\label{concl:nc-symbol}
  \sigma\big[f(\Box_A^\theta)\big](x,p)
  =f\big(
  -p^2+\Box^\theta_{A(\cdot-\frac12\Theta  p)}
  +2i\,p^\mu\,D^\theta_{A(\cdot-\frac12\Theta p)\mu}\big)1.
\end{equation}
Here, $A(\cdot-\frac12\Theta p)$ denotes the external fields $A$
shifted by the amount $\frac12\Theta p$. 

Form this formula, one might think that the new $p$-dependence in the
external fields leads to an improvement in the decay properties of the
symbol for large $p$. This point of view is however misleading when
one wants to draw conclusions for the asymptotic expansion of the
regularized trace: By a change of variables, the $p$-dependence in the
external fields disappears under the spatial integral of the
trace. This fact comes solely from the non-compactness of $\mathbbm
R^4$. It may be viewed as another manifestation of the UV/IR mixing. A
similar effect can be seen for instance in the example of an
infinitely smoothing operator on $\mathbbm R$ that has a nonvanishing
trace, see the end of Appendix \ref{App:moyal}. Therefore, on
non-compact manifolds (commutative or noncommutative), arguments
linking the asymptotic expansion of the regularized trace of an
operator to the expansion of its symbol have to be taken with
caution. For our case, we propose to use the asymptotic expansion in
$p$ of the shifted symbol
\begin{equation*}
  \sigma\big[f(\Box^\theta_A)\big](x+{\textstyle\frac12\Theta p},p)
\end{equation*}
instead. This proposal is justified rigorously by operator theoretic
arguments which show that the difference between the original operator
and a certain truncation of the asymptotic expansion of the above
shifted symbol is trace-class.
Hence, it does not contribute to the divergent part of the regularized
trace and we can safely exchange the full symbol by its truncation.
This argument can even be extended to the commutative case,
thereby proving that the coefficient of the $\log \Lambda$ part of the
regularized trace is indeed given by the (noncompact) 
Wodzicki residue. The latter
observation now can be used to explain why the expression for
$c_{\textrm{log}}$ is a gauge invariant quantity: 
Since a gauge
transformation 
conjugates the Klein-Gordon operator by some unitary operator, the
fact that $c_{\textrm{log}}$ is gauge invariant is equivalent to the
vanishing of the Wodzicki residue on commutators.

To conclude, we have seen that the methods of \psdo s are a powerful
tool for the investigation of the case studied here, yet they need
to be modified in the described way for the case of the
noncommutative Moyal plane. It would be interesting to see what
modifications are necessary to study the coupling of gravity to the
bosons through a varying matrix in $\Box_A$. This is presently under
investigation.

\section{Acknowledgements}

The authors would like to thank Edwin Langmann for posing the
questions that led to Propositions \ref{prp:comm} and \ref{prp:star},
for proposing to consider the gauge invariant expression
Eg.~(\ref{ginvdet-defn}), and for suggesting to use the representation
(\ref{Heaviside}) for the Heaviside step function. They would also
like to thank Sylvie Paycha for discussions on renormalized traces and
for providing us with Ref. \cite{Paycha:2001}, and Jouko Mickelsson for
constant interest throughout all stages of the project.\\ 
C.P. acknowledges financial support by the German Research Foundation
(DFG) under the Emmy Noether Programme.

\begin{appendix}
\section{Details of the computations}
\subsection{Computation of $c_{\textrm{log}}(A)$\label{App:clog}}
We are using the following convention for the Klein-Gordon
operator:
\begin{equation}
  \begin{split}
    \Box_A &= D_A^{\mu} D_{A \mu} = (\partial^{\mu} + ie A^{\mu}) (
    \partial_{\mu} + ie A_{\mu})\\&=   \partial^\mu
    \partial_\mu+ ie
    \partial^{\mu} A_{\mu} + 2ie A_{\mu}\partial^{\mu} - e^2 A^{\mu} A_{\mu} \\
    &=\Box_0 + ie\partial^{\mu} A_{\mu} + 2ie A_{\mu} - e^2
    A^{\mu}A_{\mu}.
  \end{split}
\end{equation}
Recall the definition of the symbol $\sigma[a]$ of a
pseudo-differential operator $a$:
\begin{equation}
  (af)(x)= \int_{\mathbbm R^4} \frac{d^4 p}{(2 \pi)^4}
  \int_{\mathbbm R^4}d^4 y e^{ip \cdot (x-y)} \sigma [a] (p,x) f(y).
\end{equation}

In the computation we need the symbol of the resolvent of the
Klein-Gordon operator, i.e. of the operator
$(c_1I+c_2\Box_A)^{-1}$. To determine an asymptotic expansion for
this symbol we start with the following expression:
\begin{equation}
  \begin{split}
    &(c_1 I + c_2 \Box_A) af(x)
    \\&= c_1 I + c_2 (\partial^\mu
    \partial_\mu + ie \partial^{\mu} A_{\mu} + 2ie A_{\mu}
    \partial^{\mu}
    \\& \quad -e^2 A^{\mu} A_{\mu}) \int_{\mathbbm R^4} \frac{d^4 p}{(2
      \pi)^4} \int_{\mathbbm R^4} d^4 y e^{ip \cdot (x-y)} \sigma [a]
    (p,x) f(y)
    \\ &=\int_{\mathbbm R^4} \frac{d^4 p}{(2 \pi)^4}
    \int_{\mathbbm R^4} d^4 y e^{ip \cdot (x-y)} (c_1 I + c_2(-p^2 -
    2e A_\mu p^\mu +
    \\ &\quad \partial^\mu
    \partial_\mu + ie \partial^\mu A_\mu + 2ie A_\mu \partial^\mu +
    2i p_\mu \partial^\mu - e^2 A^\mu A_\mu)) \sigma[a] (p,x) f(y)
  \end{split}
\end{equation}
Next replacing $a$ by $(c_1I + c_2 \Box_A)^{-1}$ we get
\begin{equation}
  \begin{split}
    &(c_1 I + c_2 \Box_A)(c_1 I + c_2 \Box_A)^{-1} af(x) = af(x)
    \\
    &= \int_{\mathbbm R^4} \frac{d^4 p}{(2 \pi)^4} \int_{\mathbbm R^4}
    d^4 y e^{ip \cdot (x-y)}(c_1 I + c_2 (-p^2 - 2eA_\mu p^\mu +
    \partial^\mu
    \partial_\mu + ie \partial^\mu A_\mu +
    \\ &\quad 2ie A_\mu \partial^\mu
    + 2i p_\mu \partial^\mu - e^2 A^\mu A_\mu)) \sigma [(c_1 I + c_2
      \Delta_A)^{-1}a] (p,x)f(y)\break
    \\ &= \int_{\mathbbm R^4}
    \frac{d^4 p}{(2 \pi)^4} \int_{\mathbbm R^4} d^4 y e^{ip \cdot
      (x-y)} \sigma[a] (p,x) f(y)
  \end{split}
\end{equation}
So we have
\begin{equation}
  \begin{split}
    &(c_1 I + c_2 (-p^2 - 2e A_\mu p^\mu +\partial^\mu
    \partial_\mu + ie \partial^\mu A_\mu + 2ie A_\mu \partial^\mu -
    e^2 A^\mu A_\mu +
    \\&2i p_\mu \partial^\mu)) \sigma[(c_1 I + c_2
      \Delta)^{-1}a] (p,x)= \sigma[a] (p,x)
  \end{split}
\end{equation}
which can be written as
\begin{equation}
  \begin{split}
    &(c_1I - c_2p^2) \sigma[(c_1 I + c_2 \Box_A)^{-1}a] (p,x) +
    \\
    &c_2( \Box_A + 2ip_\mu D^\mu_A) \sigma[(c_1 I + c_2
      \Box_A)^{-1}a] (p,x) = \sigma[a](p,x)
  \end{split}
\end{equation}
giving us the recursive relation
\begin{equation}
  \begin{split}
    &\sigma[(c_1 I + c_2 \Box_A)^{-1}a] (p,x) \\ &= \frac{1}{c_1I -
      c_2p^2} \sigma[a](p,x) - \frac{c_2}{c_1I - c_2p^2} (\Box_A
    +2ip_\mu D^\mu_A) \sigma[(c_1 I + c_2 \Box_A)^{-1}a] (p,x).
  \end{split}
\end{equation}
We can now get the desired asymptotic expansion by setting $a=1,
\sigma[a]=1$
\begin{equation}\label{resolvent}
  \sigma[(c_1 I + c_2 \Box_A)^{-1}] (p,x) \sim \sum_{n=0}^{\infty}
  \frac{(-1)^n c_2^n}{(c_1-c_2p^2)^{n+1}} (\Box_A + 2ip_\mu
  D_A^\mu)^n1.
\end{equation}

Next we evaluate explicitly the terms contributing to the
logarithmically diverging part in the expansion (\ref{trlog}) of the
effective action. When taking the angular integrals the following
formulas are used
\begin{equation*}
  \begin{split}
    &\langle p^\mu p^\nu \rangle = \frac14 p^2 \eta^{\mu \nu}
    \\&
    \langle p^{\mu_1} p^{\mu_2} p^{\mu_3} p^{\mu_4} \rangle =
    \frac{1}{24} p^4 (\eta^{\mu_1 \mu_2} \eta^{\mu_3 \mu_4} +
    \eta^{\mu_1 \mu_3} \eta^{\mu_2 \mu_4}+ \eta^{\mu_1 \mu_4}
    \eta^{\mu_2 \mu_3})
  \end{split}
\end{equation*}
where the brackets denote integration over the unit sphere in
$\mathbbm R^4$, that is
\begin{equation}
  \langle f(p) \rangle:= \frac{1}{2 \pi^2} \int_{\mathbbm R^4}
  \frac{d^4p}{(2 \pi)^4} \delta (|p|-1)f(p).
\end{equation}
Also the angular integral over an odd number of components $p^\mu$
is zero. The $s$-integrals in the expansion can be evaluated
exactly using the formula
\begin{equation}
  \int_0^1 ds \frac{s^{n-1}}{(1+sa)^{n+1}}=\frac{1}{n(1+a)^n}
\end{equation}
which holds for $\mathfrak{Re} \; a>0$. The effective action can
now be written as
\begin{equation} \label{expansion}
  \begin{split}
    &\Tr_\Lambda
    ({\textstyle \log(\frac{-\Box_A + m^2}{\Lambda_0^2})} -
    {\textstyle \log(\frac{-\Box_0 + m^2}{\Lambda_0^2})})\\
    &=- \frac{1}{(2 \pi)^4} 
    \sum_{n=1}^\infty \int_1^{\Lambda} d|p|
    |p|^3 \frac{1}{n(p^2+m^2)^n} \int_{\mathbbm R^4} d^4x \int_{S^3} d
    \Omega_p \TrN (\Box_A+2ip_\mu D_A^\mu)^n1\\[1.5ex]
    &
    \,\quad+\textrm{ const.},
  \end{split}
\end{equation}
where the constant (in $\Lambda$) term arises from the integration of
the symbol over the region $|p|\leq 1$ for which the asymptotic
expansion is not valid.

When expanding the integrand in terms of $p$, the leading term is of
the order $p^{3-2n}$ times a term of order at most $p^n$ coming from
the angular integration --- so the highest order term is of order
$p^{3-n}$. For the terms contributing to logarithmic divergences of
the effective action the leading order has to be larger than or equal
to $-1$, so the relevant terms in the expansion above are terms of
order up to four. To find the parts contributing to the logarithmic
divergence we derivate the terms in (\ref{expansion}) with respect to
$\Lambda$ and then pick the terms proportional to $1/\Lambda$. We
denote by $I_n$ the $n$th term in the expansion.  Writing the
expansions of the first four terms explicitly we then have
\begin{equation*}
  \begin{split}
    \frac{\partial I_1}{\partial \Lambda} 
    &= -
    \frac{1}{8\pi^2} \frac{\Lambda^3}{\Lambda_0^2}
    \frac{\Lambda_0^2}{\Lambda^2 + m^2} \int_{\mathbbm R^4} d^4x
    \TrN \Box_A 1 \\
    &= \frac{1}{8\pi^2} ( \Lambda - m^2
    \frac{1}{\Lambda} + O(\frac{1}{\Lambda^3})) \int_{\mathbbm
      R^4}d^4x \TrN \Box_A 1\\[2ex]
    \frac{\partial I_2}{\partial \Lambda} 
    &= - \frac{1}{8\pi^2}
    \frac{\Lambda^3}{2(\Lambda^2 + m^2)^2} \int_{\mathbbm R^4} d^4x
    \TrN \Box_A^2 1 -  \frac{1}{8\pi^2}
    \frac{\Lambda^5}{(\Lambda^2 + m^2)^2} \int_{\mathbbm R^4} d^4x
    \TrN \Box_A 1 \\ 
    &= -\frac{1}{8\pi^2}
    \frac{1}{2\Lambda} \int_{\mathbbm R^4}d^4x \TrN \Box_A^2 1
    + ... + \frac{1}{8\pi^2} m^2  \frac{1}{\Lambda} \int_{\mathbbm
      R^4}d^4x \TrN \Box_A 1 + ...
  \end{split}
\end{equation*}
\begin{equation*}
  \begin{split}
    \frac{\partial I_3}{\partial \Lambda} 
    &= -\frac{1}{8\pi^2}
    \frac{\Lambda^3}{3(\Lambda^2+m^2)^3}
    \int_{\mathbbm R^4}d^4x \mathrm{Tr_N} \Box_A^3 +
    \frac{1}{4\pi^2} \frac{\Lambda^5}{3(\Lambda^2+m^2)^3}
    \int_{\mathbbm R^4}d^4x \TrN \Box_A^2 \\
    & \quad +
    \frac{1}{8\pi^2} \frac{\Lambda^5}{3(\Lambda^2+m^2)^3}
    \int_{\mathbbm R^4}d^4x \TrN (D_A^\mu \Box_A D_{A \mu})\\
    & 
    = -\frac{1}{8\pi^2} O(\frac{1}{\Lambda^3})+...
    +\frac{1}{4\pi^2} \frac{1}{3\Lambda} \int_{\mathbbm R^4}d^4x
    \TrN \Box_A^2 + ... +\\
    &\hspace*{15em}
    + \frac{1}{8\pi^2} \frac{1}{3\Lambda}
    \int_{\mathbbm R^4}d^4x \TrN (D_A^\mu \Box_A D_{A\mu})
    \\[2ex]
    \frac{\partial I_4}{\partial \Lambda} 
    &= -
    \frac{1}{8\pi^2} \frac{\Lambda^3}{4(\Lambda^2+m^2)^4}
    \int_{\mathbbm R^4} d^4x \TrN (\Box_A^4 \\
    & \quad -
    \Lambda^2 (3\Box_A^3 + \Box_A D_A^\mu \Box_A D_{A\mu} + D_{A\mu}
    \Box_A^2 D_{A\mu}) \\
    & \quad + \frac23 \Lambda^4 (\Box_A^2 +
    D_A^\nu D_A^\mu D_{A\nu} D_{A\mu} + D_A^\mu \Box_A D_{A\mu})) \\
    &= - \frac{1}{8\pi^2} \frac23 \frac{1}{4\Lambda} \int_{\mathbbm
      R^4} d^4x \TrN (\Box_A^2 + D_A^\nu D_A^\mu D_{A\nu}
    D_{A\mu} + D_A^\mu \Box_A D_{A \mu}) + ...
  \end{split}
\end{equation*}

\subsection{Comparison with fermion calculations\label{App:Lang}}

We now compute the traces of the relevant terms in the identity
\begin{equation*}
  \begin{split}
    \log&\big(
    -\mathbbm 1_4 (\Box_A+m^2) -ie\fslash\big)\\
    &=
    \textstyle
    \log\big(\frac{\mathbbm 1_4 (-\Box_A+m^2)}{\Lambda_0^2}\big)
    +\log\big(\mathbbm 1_4-\frac{ie}{-\Box_A+m^2} \fslash\big)\\
    &\quad
    + \frac12 
    \biggl[
      \log(\mathbbm 1_4
	  {\textstyle \frac{-\Box_A+m^2}{\Lambda_0^2}}),\log
      (\mathbbm 1_4-{\textstyle \frac{ie}{-\Box_A+m^2}} \fslash) \biggr] \\
    & \quad +
    \frac{1}{12} \biggl( \left[ \log(
      \mathbbm 1_4
	       {\textstyle \frac{-\Box_A+m^2}{\Lambda_0^2}}), 
      \left[ \log(\mathbbm 1_4
	{\textstyle\frac{ -\Box_A+m^2}{\Lambda_0^2}}),
	\log (\mathbbm 1_4-{\textstyle \frac{ie}{-\Box_A+m^2}} \fslash)
	\right]\right]+ \\
    & \quad \quad 
    + \left[ \log (\mathbbm 1_4-{\textstyle\frac{ie}{-\Box_A+m^2}}
      \fslash),
      \left[ \log (\mathbbm 1_4-{\textstyle\frac{ie}{-\Box_A+m^2}}
	\fslash),
	\log(\mathbbm 1_4{\textstyle\frac{-\Box_A+m^2}{\Lambda_0^2}})
	\right] \right]
    \biggr) \\
    &\quad+ O(\textstyle \frac{1}{\Lambda^5}).
    \end{split}
\end{equation*}
The commutator terms come from the Baker-Campbell-Haus\-dorff
formula. Terms that fall off at least as $\frac1{\Lambda^{5}}$ have
been suppressed. We find
\begin{equation*}
  \begin{split}
    \Tr_\Lambda \log
    (\mathbbm 1_4({\textstyle\frac{-\Box_A+m^2}{\Lambda_0^2}}))
    &=4\Tr_\Lambda \log
    ({\textstyle\frac{-\Box_A+m^2}{\Lambda_0^2}}) \\
     \Tr_\Lambda \log
    (\mathbbm 1_4 -{\textstyle\frac{ie}{m^2-\Box_A}}\fslash) 
    &= -\int_{|p| \leq
      \Lambda}\frac{d^4p}{(2\pi)^4} \int_{\mathbbm R^4} d^4x
    \sum_{n=1}^{\infty} \frac1n \mathrm{tr}(\sigma
    [{\textstyle\frac{ie}{-\Box_A+m^2}}] \ast \fslash )^{\ast n}
  \end{split}
\end{equation*}
Here, $\ast$ denotes the product of symbols of two \psdo s which has the
asymptotic expansion (\ref{psdo-product}).

Now using the fact that $\mathrm{tr} \,\fslash = 0$ and the
expansion (\ref{resolvent}) for $\sigma [\frac{1}{c_1+c_2 \Box_A}]$,
we get
\begin{equation*}
  \begin{split}
    &\Tr_\Lambda \log (\mathbbm 1_4
    -{\textstyle \frac{ie}{m^2-\Box_A}}\fslash)\\
    &= -
    \frac12 \int_{|p| \leq \Lambda}\frac{d^4p}{(2\pi)^4}
    \int_{\mathbbm R^4} d^4x\, \mathrm{tr} (\sigma
    [{\textstyle \frac{ie}{-\Box_A+m^2}}] \ast \fslash \ast \sigma
    [{\textstyle \frac{ie}{-\Box_A+m^2}}] \ast \fslash) 
    + O(\frac{1}{\Lambda^5})\\
    &=
    \frac12 \int_{|p| \leq \Lambda}\frac{d^4p}{(2\pi)^4}
    \int_{\mathbbm R^4} d^4x  \frac{e^2}{(p^2+m^2)^2} \mathrm{tr}
    \fslash^2 + O(\frac{1}{\Lambda^5}) \\
    &= \frac{e^2}{16 \pi^2}
    \log \Lambda \int_{\mathbbm R^4} d^4x \,\mathrm{tr}
    \fslash^2 + O(\Lambda^0)
  \end{split}
\end{equation*}
This provides the results needed in the main text, since, as will be
shown below, there are no contributions to the divergent part of the
trace that come from the commutator terms. For this, we expand the
logarithm in the first commutator term above which gives us
\begin{equation*}
  \begin{split}
    &\frac12 \Tr_\Lambda 
    [\log (\mathbbm 1_4{\textstyle \frac{-\Box_A +m^2}{\Lambda_0^2}}), 
      \log (\mathbbm 1_4 -{\textstyle \frac{ie}{-\Box_A+m^2}}
      \fslash)]
    \\
    &=
    \frac12 \Tr_\Lambda [\log
      (\mathbbm 1_4
      {\textstyle \frac{-\Box_A+m^2}{\Lambda_0^2}}),
      {\textstyle \frac{-ie}{-\Box_A+m^2}}
      \fslash]
    \\
    & \quad + \frac12 \Tr_\Lambda [\log
      (\mathbbm 1_4{\textstyle \frac{(-\Box_A+m^2)}{\Lambda_0^2}}),
      -{\textstyle \frac12(\frac{ie}{-\Box_A+m^2}}
      \fslash)^2] + O({\textstyle \frac{1}{\Lambda^7}})
  \end{split}
\end{equation*}
The first term on the r.h.s is zero, so we have
\begin{equation*}
  \begin{split}
    &\frac12 \Tr_\Lambda [\log (\frac{\mathbbm 1_4 (-\Box_A +
    m^2)}{\Lambda_0^2}),
      \log (\mathbbm 1_4-\frac{ie}{-\Box_A + m^2} \fslash)] \\
    &=\frac14 e^2 \Tr_\Lambda
    [\log(\frac{\mathbbm 1_4(-\Box_A+m^2)}{\Lambda_0^2}),
      (\frac{1}{-\Box_A+m^2}
      \fslash)^2] + O(\frac{1}{\Lambda^7})
  \end{split}
\end{equation*}
Now
\begin{equation*}
  \begin{split}
    \sigma \big[
      \big[\log(\mathbbm 1_4
    {\textstyle \frac{(-\Box_A+m^2)}{\Lambda_0^2}})
    &,
    ({\textstyle \frac{1}{-\Box_A+m^2}}
    \fslash)^2\big] \big]\\
    &=\big[\log \frac{p^2+m^2}{\Lambda_0^2},(\frac{1}{p^2+m^2})^2
      \fslash^2 \big]_\ast
    + O(\frac{1}{\Lambda^6}) \\
    &= -i\frac{2p^\mu \Lambda_0^2}{p^2+m^2}(\frac{1}{p^2+m^2})^2 
    \partial_\mu
    (\fslash^2) + O(\frac{1}{\Lambda^6}) \\
    &= O(\frac{1}{\Lambda^5})
  \end{split}
\end{equation*}
so there is no contribution to the divergent part of the trace. 

Next, we turn to the first triple commutator term in the above
identity. Counting the powers of $p$ in the pertinent symbols, the
leading term should scale like $\frac1 {p^4}$. However, this term
contains a single $\fslash$ which gives zero under the trace
$\TrN$. Therefore, one has to take one more term in the expansion of
$\log(\mathbbm 1_4-\frac{ie}{-\Box_A+m^2}\fslash)$. The resulting
expression then is of order $\frac1{p^6}$ and hence can be dropped.
Finally the second triple commutator term can be seen to behave as
$\frac1{p^6}$. In conclusion, we have shown that for the divergent
terms of the cut-off regularized trace, all commutator terms can be
neglected in the above identity.

\subsection{Dependence on the regularization scheme\label{App:regdep}}

\emph{Computation of the symbol of $\theta(\Lambda+\Box_A)$.}  We
start with the following sum expression for the (regularized)
$\theta$-function \cite[p. 248 {\sl et seq.}]{FetterWalecka}
\begin{equation*}
  \theta_\epsilon (x) 
  = \frac1\epsilon
  \sum_{r=-\infty}^\infty \frac{e^{i\omega_r0^+}}{x+i\,\omega_r}
  =\frac{e^{-x0^+}}{e^{-x\epsilon}+1},\quad  \omega_r=(2r+1)\pi /\epsilon,
  \quad\epsilon>0.
\end{equation*}
(This expression is the discretized version of the well-known
integral formula 
$\theta(x)=\int_{\mathbbm R}\frac{dz}{2\pi
  i}\frac{e^{iz0^+}}{z-ix}$.
The latter is regained for $\epsilon\rightarrow \infty$.) 
Differentiation yields
\begin{equation*}
  \begin{split}
    \delta^{(n-1)}_\epsilon (x) = \frac1\epsilon
    \sum_{r=-\infty}^\infty 
    \frac{{e^{i\omega_r0^+}}(-1)^n\,n!}{(x+i\,\omega_r)^{n+1}},
    \quad
    n=1,2,3,\ldots,
  \end{split}
\end{equation*}
for the $(n-1)$th derivative of the (regularized) Dirac $\delta$-function.\\ 
Using the above expression, we have
\begin{equation*}
  \sigma [\theta_\epsilon (\Lambda^2 + \Box_A)] = 
   \frac1\epsilon
  \sum_{r=-\infty}^\infty \sigma \biggl[\frac{e^{i\omega_r0^+}}{(\Lambda^2 +
      \Box_A)+i\,\omega_r}\biggr].
\end{equation*}
We derived the asymptotic expansion for the symbol of $(c_1 I+
c_2\Box_A)^{-1}$ (see Equation (\ref{resolvent})) to be given by
\begin{equation*}
  \sigma[(c_1 I + c_2 \Box_A)^{-1}] (p,x) \sim \sum_{n=0}^{\infty}
  \frac{(-1)^n c_2^n}{(c_1-c_2p^2)^{n+1}} (\Box_A + 2ip_\mu
  D_A^\mu)^n1.
\end{equation*}
Using this we get
\begin{equation*}
  \begin{split}
    \sigma [\theta_\epsilon (\Lambda + \Box_A)]
    \sim
    \frac1\epsilon
    \sum_{r=-\infty}^\infty 
    \sum_{n=0}^{\infty} 
    \frac{{e^{i\omega_r0^+}}(-1)^n}{((\Lambda^2 - p^2)+i\omega_r)^{n+1}} 
    (\Box_A+2ip^\mu D_{A\mu})^n1
  \end{split}
\end{equation*}
Using the expressions for $\delta^{(n)}_\epsilon$ in the above
expansion we finally have
\begin{equation*}
  \sigma [\theta_\epsilon (\Lambda + \Box_A)]
  \sim\sum_{n=0}^{\infty} \frac{1}{n!}
  \delta^{(n-1)}_\epsilon(\Lambda^2 -p^2)(\Box_A+2ip^\mu D_{A\mu})^n1
\end{equation*}
\emph{Computation of the traces.}
We can now proceed with the calculation of the trace. From the
remarks in the main section, we know that we can drop the spatial
regulator $f$ since terms proportional to the volume of $\mathbbm R^4$
cancel exactly. We calculate
\begin{equation*}
  \begin{split}
    &\Tr_\Lambda^{\Box_A} \log (\frac{- \Box_0 +
      m^2}{\Lambda_0^2})-\Tr_\Lambda
    \log (\frac{- \Box_0 +
      m^2}{\Lambda_0^2})\\
    &=\sum_{n=0}^{\infty}\frac{1}{n!}
    \int_{|p|\geq 1} \frac{d^4p}{(2\pi)^4} \int_{\mathbbm R^4}d^4x
    \delta^{(n-1)}_\epsilon (\Lambda^2-p^2)
    \log({\textstyle \frac{p^2+m^2}{\Lambda_0^2}})
    \TrN(\Box_A + 2ip^\mu D_{A \mu})^n1 \\
    &\hspace*{22em} +\ldots-\Tr_\Lambda
    \log (\frac{- \Box_0 +
      m^2}{\Lambda_0^2})\\
    &= \int_{|p|\geq 1} \frac{d^4p}{(2\pi)^4}
    \int_{\mathbbm R^4} d^4x
    \theta_\epsilon(\Lambda^2-p^2)
    \log({\textstyle \frac{p^2+m^2}{\Lambda_0^2}})
    \TrN  1
    \\
    & \quad + \sum_{n=1}^{\infty}\frac{1}{n!}
    \int_{|p|\geq 1}
    \frac{d^4p}{(2\pi)^4} \int_{\mathbbm R^4} \delta^{(n-1)}_\epsilon
    (\Lambda^2-p^2)
	 {\textstyle \log(\frac{p^2+m^2}{\Lambda_0^2})}
	 \TrN (\Box_A + 2ip^\mu D_{A\mu})^n1 \\
     &\hspace*{22em}+\ldots
     -\Tr_\Lambda
     \log (\frac{- \Box_0 +
       m^2}{\Lambda_0^2}),
  \end{split}
\end{equation*}
where the dots indicate terms that are uniformly bounded in
$\Lambda$.\footnote{In particular, we have split the $p$-integral in
a part over the unit ball and an integral over the rest. The former
contributes to the finite part.} Now the first term on the r.h.s.
matches the last one in the limit $\epsilon\rightarrow \infty$. As
explained in the main text, we are interested in the terms $n\leq5$
of the sum above. Expanding the pertinent terms and performing the
angular $p$-integrals gives
\begin{equation*}
  \begin{split}
    &\Tr_\Lambda^{\Box_A} \log (\frac{- \Box_0 +
      m^2}{\Lambda_0^2})-\Tr_\Lambda
    \log (\frac{- \Box_0 +
      m^2}{\Lambda_0^2})\\
    &=\frac{1}{8\pi^2} \int_1^{\infty} dp\,p^3
    \delta(\Lambda^2-p^2)
    {\textstyle \log (\frac{p^2+m^2}{\Lambda_0^2})}
    \int_{\mathbbm R^4} d^4x \,\TrN \Box_A1 \\
    & \quad
    + \frac12 \frac{1}{8\pi^2}
    \int_1^{\infty} {dp}\,p^3 \delta^{(1)}(\Lambda^2-p^2)
    {\textstyle \log(\frac{p^2+m^2}{\Lambda_0^2})}
    \int_{\mathbbm R^4}d^4x \,\TrN (\Box_A^2 -p^2 \Box_A)1 \\
    & \quad 
    + \frac{1}{6}
    \frac{1}{8\pi^2} \int_1^{\infty} dp\, p^3
    \delta^{(2)}(\Lambda^2-p^2)
    {\textstyle \log(\frac{p^2+m^2}{\Lambda_0^2})}
    \int_{\mathbbm R^4}d^4x\,
    \TrN (\Box_A^3 -p^2(2 \Box_A^2 +\\
    &\hspace{25em}+ D_{A \mu}
    \Box_A D_{A\mu}))1 \\
    & \quad + \frac{1}{24}\frac{1}{8\pi^2} \int_1^{\infty}
    dp\, p^3 \delta^{(3)}(\Lambda^2-p^2)
    {\textstyle \log(\frac{p^2+m^2}{\Lambda_0^2})}
    \int_{\mathbbm R^4}d^4x \,
    \TrN \bigg(\Box_A^4 -3p^2 \Box_A^3 -\\
    &\hspace*{5em}
    -p^2 \Box_A D_{A\mu}
    \Box_A  D_A^\mu - p^2 D_{A \mu} \Box_A^2 D_{A \mu} -p^2
    D_{A\mu} \Box_A D_A^\mu \Box_A+
    \\[1ex]
    &\hspace*{10em}
    + {\textstyle\frac23} p^4(\Box_A^2 + D_A^\mu
    D_A^\nu D_{A \mu} D_{A \nu} + D_A^\mu \Box_A D_{A \mu})\bigg)1 +\ldots
  \end{split}
\end{equation*}
We have also taken the limit $\epsilon\rightarrow \infty$, in which
$\delta_\epsilon$ goes over into the Dirac $\delta$-function.\\
Gathering terms with equal spatial integral we obtain
\begin{equation*}
  \begin{split}
    &\Tr_\Lambda^{\Box_A} \log (\frac{- \Box_0 +
      m^2}{\Lambda_0^2}) - \Tr_\Lambda^{\Box_0} \log (\frac{- \Box_0
      + m^2}{\Lambda_0^2}) \\
    &=\frac{1}{8\pi^2} \int_1^{\infty}
    {dp}\log ({\textstyle \frac{p^2 + m^2}{\Lambda_0^2}}) \biggl(p^3
    \delta(\Lambda^2-p^2)
    - {\textstyle\frac12} p^5 \delta^{(1)}(\Lambda^2 -
    p^2) \biggr)
    \int_{\mathbbm R^4}d^4x \,\TrN \Box_A1\\
    & \quad
    +\frac{1}{8\pi^2}
    \int_1^{\infty} {dp}
    \log ({\textstyle \frac{p^2 + m^2}{\Lambda_0^2}})
    \biggl( {\textstyle\frac12} p^3 \delta^{(1)}(\Lambda^2-p^2)
    - {\textstyle\frac13} p^5 \delta^{(2)}(\Lambda^2 - p^2) \\
    & \hspace*{16em}
    + {\textstyle\frac{1}{36}} p^7 \delta^{(3)}(\Lambda^2 - p^2) \biggr)
    \int_{\mathbbm R^4}d^4x\, \TrN \Box_A^21 \\
    & \quad +
    \frac{1}{8\pi^2} \int_1^{\infty} {dp}\log (\frac{p^2 +
      m^2}{\Lambda_0^2}) \biggl(- {\textstyle\frac16} p^5
    \delta^{(2)}(\Lambda^2-p^2) \\
    &  \hspace*{13em}
    + {\textstyle\frac{1}{36}} p^7
    \delta^{(3)}(\Lambda^2 - p^2) \biggr)
    \int_{\mathbbm R^4}d^4x\, \TrN  D_A^\mu \Box_A D_{A \mu}1 \\
    & \quad + \frac{1}{8\pi^2}
    \int_1^{\infty} {dp}\log ({\textstyle\frac{p^2 + m^2}{\Lambda_0^2}})
    {\textstyle\frac{1}{36}} p^7 \delta^{(3)}(\Lambda^2-p^2)
    \int_{\mathbbm R^4}d^4x
    \TrN  D_A^\mu D_A^\nu D_{A \mu} D_{A \nu}1 \\
    & \quad
    + \frac{1}{8\pi^2} \int_1^{\infty} {dp}
    \log ({\textstyle\frac{p^2 + m^2}{\Lambda_0^2}})
    \biggl({\textstyle\frac16} p^3 \delta^{(3)}(\Lambda^2-p^2)\\
    &\hspace*{15em}
    - {\textstyle\frac18} p^5 \delta^{(4)}(\Lambda^2 - p^2)\biggr)
    \int_{\mathbbm R^4}d^4x \,\TrN \Box_A^31
    + \ldots
  \end{split}
\end{equation*}
Next we make a change of variables $p^2=u$ to get
\begin{equation*}
  \begin{split}
    &\Tr_\Lambda^{\Box_A} 
    \log (\frac{- \Box_0 + m^2}{\Lambda_0^2}) 
    - \Tr_\Lambda^{\Box_0} \log (\frac{-\Box_0 + m^2}{\Lambda_0^2}) 
    \\
    &=\frac{1}{8\pi^2} \int_1^{\infty}
    \frac{du}{2}{\textstyle\log (\frac{u + m^2}{\Lambda_0^2})} \biggl(u
    \delta(\Lambda^2-u)   - {\textstyle\frac12} u^2
    \delta^{(1)}(\Lambda^2 -
    u) \biggr) \int_{\mathbbm R^4}d^4x \,\TrN \Box_A1 
    \\
    & \quad
    +\frac{1}{8\pi^2} \int_1^{\infty} \frac{du}{2} 
    {\textstyle\log (\frac{u + m^2}{\Lambda_0^2})}
    \biggl( {\textstyle\frac12} u \delta^{(1)}(\Lambda^2-u)
    - {\textstyle\frac13} u^2 \delta^{(2)}(\Lambda^2 - u) 
    \\
    & \hspace*{16em}
    +{\textstyle \frac{1}{36}} 
    u^3 \delta^{(3)}(\Lambda^2 - u) \biggr)
    \int_{\mathbbm R^4}d^4x \,\TrN \Box_A^21+
  \end{split}
\end{equation*}
\begin{equation*}
  \begin{split}
    & \quad +
    \frac{1}{8\pi^2} \int_{1}^{\infty} 
    \frac{du}{2} {\textstyle\log (\frac{u + m^2}{\Lambda_0^2})} 
    \biggl(- {\textstyle\frac16} u^2
    \delta^{(2)}(\Lambda^2-u)
    + {\textstyle\frac{1}{36}} u^3 \delta^{(3)}(\Lambda^2 - u)
    \biggr)\times\\
    &\hspace*{21em}\times
    \int_{\mathbbm R^4}d^4x
    \TrN  D_A^\mu \Box_A D_{A\mu}1 
    \\
    & \quad + \frac{1}{8\pi^2} \int_1^{\infty} \frac{du}{2}
    \log {\textstyle (\frac{u + m^2}{\Lambda_0^2})}
    {\textstyle \frac{1}{36}} u^3
    \delta^{(3)}(\Lambda^2-u) \int_{\mathbbm R^4}d^4x
    \TrN D_A^\mu D_A^\nu D_{A \mu} D_{A \nu}1
    \\
    & \quad +
    \frac{1}{8\pi^2} \int_0^{\infty} \frac{du}{2}
    {\textstyle \log (\frac{u + m^2}{\Lambda_0^2})}
    \biggl({\textstyle \frac16} u \delta^{(2)}(\Lambda^2-u) -
    {\textstyle \frac18} u^2 \delta^{(3)}(\Lambda^2 - u)\biggr)
    \times\\
    &\hspace*{22em}\times
    \int_{\mathbbm R^4}d^4x \,\TrN \Box_A^31 + \ldots
  \end{split}
\end{equation*}
Now by integrating by parts and noting that
\begin{equation*}
  \frac{d^k}{du^k}
  \delta(\Lambda^2-u)=(-1)^k\delta^{(k)}(\Lambda^2-u)
\end{equation*}
we have
\begin{equation*}
  \begin{split}
    &\Tr_\Lambda^{\Box_A} \log (\frac{- \Box_0 +
      m^2}{\Lambda_0^2})
    - \Tr_\Lambda^{\Box_0} \log (\frac{-\Box_0
      + m^2}{\Lambda_0^2}) \\
    &=\frac{1}{16\pi^2} \int_1^{\infty} du
    {\textstyle \log  (\frac{u + m^2}{\Lambda_0^2})}
    \delta(\Lambda^2-u) u (1-1)
    \int_{\mathbbm R^4}d^4x \,\TrN \Box_A1 \\
    & \quad +
    \frac{1}{16\pi^2} \int_1^{\infty} du
    {\textstyle \log (\frac{u +m^2}{\Lambda_0^2})}
    \delta(\Lambda^2-u)
    (\frac12-\frac23+\frac16)
    \int_{\mathbbm R^4}d^4x \, \TrN \Box_A^21 \\
    & \quad + \frac{1}{16\pi^2} \int_1^{\infty}
    du
    {\textstyle \log (\frac{u + m^2}{\Lambda_0^2})}
    \delta(\Lambda^2-u)
    (-\frac13+\frac16)
    \int_{\mathbbm R^4}d^4x \,\TrN  D_A^\mu \Box_A D_{A\mu}1 \\
    & \quad + \frac{1}{16\pi^2} \int_1^{\infty} du
    {\textstyle \log (\frac{u + m^2}{\Lambda_0^2})} \frac16
    \delta(\Lambda^2-u)
    \int_{\mathbbm R^4}d^4x
    \TrN  D_A^\mu D_A^\nu D_{A\mu} D_{A\nu}1 \\
    &\quad+\ldots\\
    \quad&=-\frac{1}{16\pi^2} \frac16 \int_1^{\infty} du
    {\textstyle \log (\frac{u + m^2}{\Lambda_0^2})}
    \delta(\Lambda^2-u) \int_{\mathbbm R^4}d^4x
    \Tr D_A^\mu \Box_A D_{A\mu}1 \\
    & \quad
    + \frac{1}{16\pi^2}
    \frac16 \int_1^{\infty} du
    {\textstyle \log (\frac{u + m^2}{\Lambda_0^2})}
    \delta(\Lambda^2-u)
    \int_{\mathbbm R^4}d^4x \,\TrN  D_A^\mu D_A^\nu D_{A\mu} D_{A\nu}1 \\
    &\quad+\ldots\\
    &=-\frac{1}{96\pi^2}
    {\textstyle \log(\frac{\Lambda^2 + m^2}{\Lambda_0^2})}
    \int_{\mathbbm R^4}d^4x
    \TrN  D_A^\mu \Box_A D_{A\mu}1  \\
    & \quad +
    \frac{1}{96\pi^2}
    {\textstyle \log (\frac{\Lambda^2 + m^2}{\Lambda_0^2})}
    \int_{\mathbbm R^4}d^4x \,\TrN  D_A^\mu D_A^\nu D_{A\mu}
    D_{A\nu}1\\
    &\quad+\ldots
  \end{split}
\end{equation*}
Recalling that
\begin{equation*}
  \TrN  (D_A^\mu \Box_A D_{A\mu}-D_A^\nu D_A^\mu D_{A\nu}
  D_{A\mu})=\frac{e^2}2 \TrN  F^{\mu \nu} F_{\mu \nu}
\end{equation*}
we finally get
\begin{equation}
  \begin{split}
    \Tr_\Lambda^{\Box_A} \log (\frac{- \Box_0 +
      m^2}{\Lambda_0^2}) &- \Tr_\Lambda^{\Box_0} \log (\frac{-\Box_0
      + m^2}{\Lambda_0^2})\\
    &=-\frac12 \frac{1}{96 \pi^2} \log\frac{\Lambda}{\Lambda_0}
    \int_{\mathbbm R^4}\TrN
    F^{\mu \nu} F_{\mu \nu}+\ldots,
  \end{split}
\end{equation}
where again the dots indicate terms that are bounded or polynomial
in $\Lambda$.

\subsection{Computation on the Moyal plane\label{App:moyal}}

\emph{General remarks.}
The symbol of the operator $c_1+c_2\Box_A^\theta$ is given by
\begin{equation*}
  \begin{split}
    \sigma(x,p)&:=\sigma[(c_1+c_2\Box_A^\theta)^{-1}](x,p)
    \\
    &\textstyle
    \;=-p^2-2e p^\mu A_\mu(x-\frac12\Theta p)
    +i e (\partial^\mu A_\mu)(x-\frac12\Theta p)\\
    &\hspace*{12em}
    \textstyle
    -e^2(A^\mu\star A_\mu)(x-\frac12\Theta p).
  \end{split}
\end{equation*}
From this expression, it is clear that one can bound $\sigma$ from
below by a positive constant and from above by a multiple of $p^2$ for
$p^2$ greater than a certain constant.  Furthermore, the derivatives
of $\sigma$ fall off as long as $x$ is confined to some compact
set. Therefore, by \cite[corollary 5.1]{Shubin}, there is a \psdo\
that inverts $(c_1+c_2\Box_A^\theta)$ up to some infinitely smoothing
operator.\\
\emph{Derivation of the recursion relation.}
As in Section (\ref{App:clog}), we start with the following identity
for $\sigma$:
\begin{equation*}
\begin{split}
\psi(x)&=(c_1+c_2\Box_A^\theta)(c_1+c_2\Box_A^\theta)^{-1}\psi(x)\\
&=
(c_1+c_2\Box_A^\theta
\int{\textstyle\frac {d^4p}{(2\pi)^4}}
\int d^4y\,
e^{ip(x-y)}
\sigma(x,p)\,\psi(y)\\
& =
\int {\textstyle \frac{d^4p}{(2\pi)^4}}
\int d^4y\,
(c_1 +c_2(\partial^\mu \partial_\mu
+ ie(\partial^\mu A_\mu) \star + 2ieA^\mu \star
\partial_\mu \\
& \quad -e^2 A^\mu \star A_\mu \star))(e^{ip(x-y)}
\sigma[(c_1+c_2\Box_A^\theta)^{-1}](x,p))\psi(y)
\end{split}
\end{equation*}
To continue we need the following formula
\begin{equation*}
e^{ip(x-y)} \sigma(x,p) = [\sigma(\cdot+\frac12 \Theta p,p) \star
e^{ip(\cdot-y)} \chi(\cdot)](x)
\end{equation*}
which can be proved as follows. Using the integral expression for
the star product
\begin{equation*}
(f \star g)(x) := (2\pi)^{-4} \int \int e^{i\xi(x-y)}
f(x-\frac12\Theta\xi)g(y)d^4yd^4\xi
\end{equation*}
we have for a Schwartz test function $\chi$
\begin{equation*}
\begin{split}
&[\sigma(\cdot+\frac12 \Theta p,p)
\star e^{ip(\cdot-y)}\chi(\cdot)](x)\\
&=\frac{1}{(2\pi)^4}
\int\int d^4\xi
d^4z\,\sigma(x-\frac12\theta\xi+\frac12\Theta p,p)e^{ip(z-y)}\chi
(z) e^{i \xi (x-z)} \\
&=\frac{1}{(2\pi)^4}\int\int
d^4\xi d^4z\,\sigma(x-\frac12\theta\xi+\frac12\Theta p,p)\chi (z)
e^{-iz(\xi-p)}e^{i(\xi x-py)} \\
&=\frac{1}{(2\pi)^{2}}
\int d^4 \xi\,
\sigma(x-\frac12\Theta(\xi-p),p)\hat{\chi} (\xi-p)e^{i(\xi x-py)}
\\
&=e^{ip(x-y)}\frac{1}{(2\pi)^{2}}\int d^4 \tilde{\xi}\,
\sigma(x-\frac12\Theta \tilde{\xi},p)\hat{\chi} (\tilde{\xi})e^{i
\tilde{\xi} x}.
\end{split}
\end{equation*}
Now in the limit $\chi\rightarrow 1$, the Fourier transform
$\hat\chi$ approximates the delta function. Therefore, in this
limit, we obtain the claimed identity.  Using this formula in the
expression for $\psi(x)$ we get
\begin{equation*}
\begin{split}
&\psi(x) = \int \int \frac{d^4p d^4y}{(2\pi)^4} (c_1 +
c_2(\partial^\mu
\partial_\mu + ie(\partial^\mu A_\mu) \star + 2ieA^\mu \star
\partial_\mu
\\
&\hspace*{15em} -e^2 A^\mu \star A_\mu
\star)(\sigma(x,p)e^{ip(x-y)})\psi(y) \\
& =\int \int \frac{d^4p \,d^4y}{(2\pi)^4}
(c_1 + c_2(\partial^\mu
\partial_\mu + ie(\partial^\mu A_\mu) \star + 2ieA^\mu \star
\partial_\mu -e^2 A^\mu \star A_\mu
\star) \times
\\
&\hspace*{15em}  \times (\sigma(\cdot+\frac12 \Theta p,p)\star
e^{ip(\cdot-y)})(x)\psi(y) \\
& =\int \int
\frac{d^4p\,d^4y}{(2\pi)^4} [(c_1 + c_2(\Box_0 + 2ip^\mu
\partial_\mu - p^2 + ie(\partial^\mu A_\mu) \star \\
& \quad - 2ep^\mu A_\mu \star + 2ieA^\mu \star
\partial_\mu -e^2 A^\mu \star A_\mu
\star)\sigma(\cdot+\frac12 \Theta p,p)]\star
e^{ip(\cdot-y)}(x)\psi(y) \\
& =\int \int \frac{d^4p\,d^4y}{(2\pi)^4}
(c_1 +c_2p^2
+c_2(\Box_{A(\cdot-\frac12\Theta p)}^\theta
+ 2p^\mu D_{A(\cdot-\frac12\Theta p),\mu}^\theta))
\sigma(\cdot,p)e^{ip(x-y)}\psi(y)
\end{split}
\end{equation*}
which gives us
\begin{equation*}
1= (c_1 - c_2 p^2 + c_2(\Box_{A(\cdot-\frac12\Theta p)}^\theta +
2ip^\mu D_{A(\cdot-\frac12\Theta p),\mu}^\theta))\sigma [(c_1+c_2
\Box_A^\theta)^{-1}](x,p)
\end{equation*}
or
\begin{equation}\label{App:recursion}
\begin{split}
\sigma[&(c_1+c_2\Box_A^\theta)^{-1}](x,p)
\\
&\textstyle=\frac1{c_1-c_2p^2}
-
\frac {c_2}{c_1-c_2p^2}
\left(\Box^\theta_{A(\cdot-\frac12\Theta p)}
+2 i\,p^\mu\,D^\theta_{A(\cdot-\frac12\Theta p)}\right)
\sigma[(c_1+c_2\Box_A^\theta)^{-1}](x,p).
\end{split}
\end{equation}

\noindent
\emph{Derivation of the asymptotic expansion.}
We set $R:=(c_1+c_2\Box_A^\theta)^{-1}$. As $-\Box_A^\theta$ is a positive
operator, $R$ is bounded for $c_1\cdot c_2<0$. Indeed, from
\begin{equation*}
\int_{\mathbbm R^4}d^4x\,\overline{\psi}(x)\left(A\star\varphi\right)(x)
=
\int_{\mathbbm R^4}d^4x\,\left(\overline{\psi}\star A\star\varphi\right)(x)
=
\int_{\mathbbm R^4}d^4x\,\left(\overline{\psi}\star A\right)(x)
\varphi(x)
\end{equation*}
which holds for $\psi,A,\varphi\in L^2(\mathbbm R^4)$
\cite[lemma 2.10]{SpectralTriples} and
$\overline{\overline {A}\star\psi}=\overline{\psi}\star A$ we conclude
\begin{equation*}
  \langle \psi, A\star \varphi\rangle
  =\langle \overline A\star\psi,\varphi\rangle
\end{equation*}
and hence $(D_{A\mu}^\theta)^\dagger=-D_{A\mu}^\theta$. Therefore,
\begin{equation*}
  \langle \varphi,-\Box_A^\theta\varphi\rangle
  =\sum_{\mu=1}^4
  \langle D_{A\mu}^\theta\varphi, D_{A\mu}^\theta\varphi\rangle\geq0.
\end{equation*}
In our case, we have $c_1=1-s+s\frac{m^2}{\Lambda^2_0}$
and $c_2=-\frac s{\Lambda_0^2}$ for $0\leq s\leq 1$ which meets the
above requirement of
$c_1\cdot c_2<0$ for $0<s\leq 1$. For $s=0$, we have $c_2=0$, $c_1\neq
0$, and $R$ is a multiple of the identity. 

Next, let $R_N$ be the \psdo\ defined by the symbol
\begin{equation*}
  \sigma[R_N](x,p)=
  \sum_{n=0}^N
  \frac{(-c_2)^n}{(c_1-c_2p^2)^{n+1}}
  \left(\Box_{A(\cdot-\frac12\Theta p)}^\theta
  +2ip^\mu D_{A(\cdot-\frac12\Theta p),\mu}^\theta\right)^n1.
\end{equation*}
We will show that the difference $R-R_N$ is a trace class operator.\\
For this, we first apply $c_1+c_2\Box_A^\theta$ from the left to obtain
\begin{equation*}
(c_1+c_2\Box_A^\theta)(R-R_N)=1-(c_1+c_2\Box_A^\theta)R_N
\end{equation*}
Here, $1$ denotes the identity operator.  We will compute the symbol
of the \psdo\ on the rhs. of this equation.  On the level of symbols,
multiplication of $R_N$ by $c_1+c_2\Box_A^\theta$ from the left
amounts to the application of
$c_1+c_2(-p^2+\Box_{A(\cdot-\frac12\Theta p)}^\theta +2ip^\mu
D_{A(\cdot-\frac12\Theta p),\mu}^\theta)$ to $\sigma[R_N]$, cf. the
derivation of the recursion relation above.  Hence, we find
\begin{equation*}
  \begin{split}
    \sigma[1-(c_1+&c_2\Box_A^\theta)R_N](x,p)=\\
    &=1-(c_1+c_2(-p^2+\Box_{A(\cdot-\frac12\Theta p)}^\theta
    +2ip^\mu D_{A(\cdot-\frac12\Theta p),\mu}^\theta)
    \sigma[R_N](x,p)\\
    &=1-(c_1-c_2p^2)\sigma[R_N]-c_2(\Box_{A(\cdot-\frac12\Theta p)}^\theta
    +2ip^\mu D_{A(\cdot-\frac12\Theta p),\mu}^\theta)\sigma[R_N]\\
    &=-\sum_{n=1}^N
    \frac{(-c_2)^n}{(c_1-c_2p^2)^n}
    \left(\Box_{A(\cdot-\frac12\Theta p)}^\theta
    +2ip^\mu D_{A(\cdot-\frac12\Theta p),\mu}^\theta\right)^n1\\
    &\quad +
    \sum_{n=0}^N
    \frac{(-c_2)^{n+1}}{(c_1-c_2p^2)^{n+1}}
    \left(\Box_{A(\cdot-\frac{1}{2}\Theta p)}^\theta
    +2ip^\mu D_{A(\cdot-\frac12\Theta p),\mu}^\theta\right)^{n+1}1\\
    &= \frac{(-c_2)^{N+1}}{(c_1-c_2p^2)^{N+1}}
    \left(\Box_{A(\cdot-\frac12\Theta p)}^\theta
    +2ip^\mu D_{A(\cdot-\frac12\Theta p),\mu}^\theta\right)^{N+1}1.
  \end{split}
\end{equation*}
Let $r_N$ be defined by the last expression,
\begin{equation*}
  \sigma[r_N](x,p):= \frac{(-c_2)^{N+1}}{(c_1-c_2p^2)^{N+1}}
    \left(\Box_{A(\cdot-\frac12\Theta p)}^\theta
    +2ip^\mu D_{A(\cdot-\frac12\Theta p),\mu}^\theta\right)^{N+1}1.
\end{equation*}
We will show that $r_N$ is a trace class operator for sufficiently large $N$.
Expanding the power of operators in the symbol $\sigma[r_N]$ yields terms of
the form
\begin{equation*}
\textrm{const. }\times \frac1{(c_1-c_2p^2)^{N+1}}
\times f_1\star\ldots\star f_k(x-\frac12\Theta p),
\end{equation*}
$k=1,\ldots,2(N+1)$, the $f_i$ denoting the external fields $A_\mu$ or
derivatives thereof.  (We have used the fact that
$\left(f(\cdot-\frac12\Theta p)\star g(\cdot-\frac12\Theta
p)\right)(x) =(f\star g)(x-\frac12\Theta p)$.)

As $A_\mu$ is in $\mathcal P$ and of order $-2-\epsilon$, Moyal
multiplication by it increases the decay property of the $x$-dependent
part by two.  On the other hand, differentiation increases it only by
one.  Therefore, the leading term of the above type will be the one
where $N$ derivatives of $D_{A(\cdot-\frac12\Theta p),\mu}^\theta$ hit
a single $A_\mu$.  The resulting term can be bounded from above by
\begin{equation*}
\textrm{const. }\times\frac1{(1+p^2)^{N+1}}\times (p^2)^{N/2}\times
\frac1{(1+(x-\frac12\Theta p)^2)^{\frac{4+\epsilon+N}2}}
\end{equation*}
which is integrable in $x-p$ space for sufficiently large $N$.

Application from the left of the bounded operator $R$ to $r_N$ does not
change the property
of being trace class. On the other hand, we find
\begin{equation*}
R r_N=R(c_1+c_2\Box_A^\theta)(R-R_N)=R-R_N.
\end{equation*}
To summarize: If we are interested in the singular behavior of the
cut-off regularized trace of $R$, we may use the symbol
$\sigma[R_N]$ for $N$ sufficiently large in the integral formula of
the trace. This amounts to the iteration of the recursion relation
(\ref{App:recursion}) $N$ times.

\noindent\emph{Remarks.} It is easy to see that a blind application of
the machinery of \psdo\ leads astray. As already mentioned in the main
text, the symbol of the operator $f\star$ is given by
\begin{equation*}
\sigma[f\star](x,p)=f(x-{\textstyle\frac12}\Theta p).
\end{equation*}
Hence, $f\star$ is an infinitely smoothing operator if $f$ is a
Schwartz test function.  In other words, the noncommutative
Klein-Gordon operator $\Box_A^\theta$ differs from the free operator
$\Box_0$ by an infinitely smoothing operator,
\begin{equation*}
\begin{split}
\textstyle
\sigma[\Box_A^\theta](x,p)
&=
\textstyle
-p^2-2ep^\mu A_\mu(x-\frac12\Theta p)
+ie(\partial^\mu A_\mu)(x-\frac12 \Theta p)
\\
&\hspace*{15em}
\textstyle
-e^2(A^\mu\star A_\mu)(x-\frac12\Theta p)\\
&=\sigma[\Box_0](x,p)+\textrm{ smoothing.}
\end{split}
\end{equation*}
One might therefore expect that the dependence on the fields $A$ of
the resolvent $R$ is in the part that is not seen by an asymptotic
expansion in $p$ and hence does not contribute to the divergent
behavior of the trace. For \psdo s on \emph{non-compact} manifolds
$M$ this line of reasoning has to be taken with caution, since there
might be additional divergent terms from the $x$-integration in the
trace integral.  This is nicely illustrated by the above computation
and the following example.  Consider the function
\begin{equation*}
f(x,p)=e^{-x^2 \,e^{-p^2}-\frac14 p^2},
\end{equation*}
where $x$ and $p$ are one-dimensional variables. Clearly
\begin{equation*}
  |\partial^\alpha_p\partial^\beta_x f(x,p)|\leq
  C_{K,\alpha,\beta}e^{-\frac14p^2},\quad
  x\in K\subset\mathbbm R\textrm{ compact},\;p \in\mathbbm R,
\end{equation*}
hence $f$ defines an infinitely smoothing operator. On the other hand
\begin{equation*}
\int_{\mathbbm R}dx \,f(x,p)=\sqrt{\pi}e^{\frac14 p^2},
\end{equation*}
and the operator $f$ does have a diverging trace. Note that in this
example, it is the noncompactness that yields the surprise. We
conclude that even in the commutative case, the correspondence between
the logarithmically divergent part of the trace and the residue needs
some additional justification.\\
In the above calculation, however, the $p$-$x$ mixing in the arguments
of the fields $A_\mu$ --- which originates from the noncommutativity
of the Moyal plane --- makes it impossible to distinguish between the
asymptotic $p$-expansion and an (infinitely smoothing) remainder.
There, additional arguments are imperative. Observe, however, that our
lines of reasoning above can be taken over to the commutative case,
thereby solving the raised objection.

\end{appendix}

\end{document}